\documentclass[pre,aps,singlecolumn,superscriptaddress,preprintnumbers,nopacs,tightenlines,floatfix,amsmath,amssymb]{revtex4}

\usepackage{bm}

\usepackage{amsmath}
\usepackage{graphicx}
\usepackage{amssymb}
\usepackage{mathrsfs}
\usepackage{bbm}

\newcommand{\la}[1]{\label{#1}}
\newsavebox{\marginbox}

\newenvironment{comment}{%
  \begin{lrbox}{\marginbox}
  \begin{minipage}{\marginparwidth}
    \footnotesize
}{
  \end{minipage}
  \end{lrbox}
  \marginpar{\usebox{\marginbox}}
}

\makeatother

\usepackage{lipsum}

\newcommand{\be}{\begin{eqnarray}}
\newcommand{\ee}{\end{eqnarray}}

\begin{document}

\hskip .0cm
\title{{\large \bf  The Discrete Frenet  Frame, Inflection Point Solitons \\
 And Curve Visualization
with Applications to Folded Proteins  }}

\author{Shuangwei Hu}
\affiliation{Department of Physics and Astronomy, Uppsala University,
P.O. Box 803, S-75108, Uppsala, Sweden}
\affiliation{
Laboratoire de Mathematiques et Physique Theorique
CNRS UMR 6083, F\'ed\'eration Denis Poisson, Universit\'e de Tours,
Parc de Grandmont, F37200, Tours, France}
\author{Martin Lundgren}
\affiliation{Department of Physics and Astronomy, Uppsala University,
P.O. Box 803, S-75108, Uppsala, Sweden}
\author{Antti J. Niemi}
\affiliation{Department of Physics and Astronomy, Uppsala University,
P.O. Box 803, S-75108, Uppsala, Sweden}
\affiliation{
Laboratoire de Mathematiques et Physique Theorique
CNRS UMR 6083, F\'ed\'eration Denis Poisson, Universit\'e de Tours,
Parc de Grandmont, F37200, Tours, France}


\begin{abstract}
We develop a transfer matrix formalism  to visualize the framing of discrete piecewise linear curves in three
dimensional space. Our approach  is based on the concept of an 
intrinsically discrete curve, which enables us to more effectively describe curves that in the limit where the 
length of line segments vanishes approach fractal structures   in lieu of  continuous curves. We  verify that in the case of differentiable
curves the continuum limit of our discrete equation does reproduce the generalized Frenet  equation.  As 
an application we consider folded proteins, their Hausdorff dimension is known to be fractal.  
We explain how to employ the orientation  of  $C_\beta$ carbons of amino acids  along a protein backbone 
to introduce a preferred framing along the backbone.  By analyzing the experimentally resolved fold 
geometries in the Protein Data Bank  we observe that this  $C_\beta$ framing relates intimately to the
discrete Frenet framing. We also explain how inflection points can be  located in the loops, and clarify their
distinctive r\^ole in determining the  loop structure  of foldel proteins.
\end{abstract}

\date{\today}

\maketitle
\section{ {\bf I:} introduction}

The visualization of a three dimensional discrete framed curve is an important and widely studied topic in computer graphics, 
from the association of ribbons and tubes  to the determination of camera gaze directions along trajectories.  
Potential applications range from aircraft and robot kinematics to stereo reconstruction and 
virtual reality \cite{hansonbook},  \cite{kuipers}. 

We are interested in addressing  the problem of  characterizing the physical laws that govern protein folding.  
For this we develop a technique for  framing a general discrete and piecewise linear curve in a manner that 
will eventually enable us to combine the geometric problem of framing with an appropriate physical  principle
for frame determination. 
Our ultimate goal is to have an approach, where instead of purely geometric considerations
the frames along  a curve are determined directly from the properties of an underlying physical system. 
As a consequence we expect  that our formalism and our results will  find wide applicability well beyond the 
protein folding problem.

The classical theory of  continuous curves in three dimensional space employs the Frenet equation \cite{hansonbook},  \cite{kuipers}  to determine
a moving coordinate frame along a sufficiently differentiable space curve.  However, 
if the curve has inflection points and/or straight segments or if it  fails to be at least three times  continuously differentiable,
the Frenet frame becomes either discontinuous or may not even exist.  In such cases there can  be good reasons
to consider the option to introduce an alternative framing such as Bishop's parallel transport frame \cite{bishop},  
a geodetic reference frame  or some possibly hybrid variants  \cite{hansonbook}, \cite{kuipers}.

In this article we derive a discrete version of the Frenet equation  that introduces a framing 
along an intrinsically discrete and piecewise linear curve in $\mathbb R^3$. We develop the general formalism  for the visualization
of  such a curve without any underlying assumption that it approaches a   
continuous space curve in the limit where the maximum length of its  line 
segments goes to zero.  The continuum limit may as well be
a fractal,  with a nontrivial Hausdorff dimension. Thus, unlike in several approaches that we are aware of, 
our starting point is not  in  a discretization of the continuum Frenet equation. Instead our approach is 
intrinsically discrete,  and it is based on the transfer matrix formalism that is widely used for example in 
lattice field theories \cite{lattice}. Indeed, we find it useful to adapt some notions of lattice gauge theories \cite{lattice}.
For us this provides a valuable conceptual point of view.
Moreover, since the transfer matrix formalism intrinsically incorporates self-similarity and the very concept of line segment length 
has no r\^ole in our derivations, we can effortlessly  consider curves that  have fractal continuum limits 
while at the same time ensuring that if the continuum  limit exists as a class $\mathcal C^3$  
space curve we recover the standard Frenet framing together with  its generalized versions. 

As an application we consider folded proteins, for which the continuum limit is known to be a fractal with Hausdorff dimension that is very close
to three \cite{ulf}.  The locations of the central $C_\alpha$ carbon atoms along the protein
determines a discrete piecewise linear curve, this is  the protein backbone.  We introduce a framing to the backbone
by employing the $C_\beta$ carbon atoms of  the side chain amino acids that are covalently bonded to the
$C_\alpha$ carbons that define the backbone. The frame at the location of a given $C_\alpha$ carbon is determined by the directional vector that connects 
it  with the ensuing  $C_\beta$ carbon, together with the directional vector that
connects it to the next $C_\alpha$ carbon along the backbone.  By inspecting the framing of all  protein structures  in 
the Protein Data Bank (PDB) \cite{pdb} we  find that such a $C_\beta$ framing relates intimately to  the discrete 
Frenet framing of the backbone.  In particular, we conclude that for a folded protein the concept of an inflection point acquires an intrinsic 
biological interpretation, it coincides with the location of the center of the loop:  The inflection points  drive the protein loop geometry.

At an isolated inflection point of a continuous curve, the  curvature  which is a frame {\it independent} geometric characteristic of the curve vanishes.
At such a point the Frenet frame can become discontinuous (see Figure 1). 
\begin{figure}[h]
        \centering
                \includegraphics[width=0.4\textwidth]{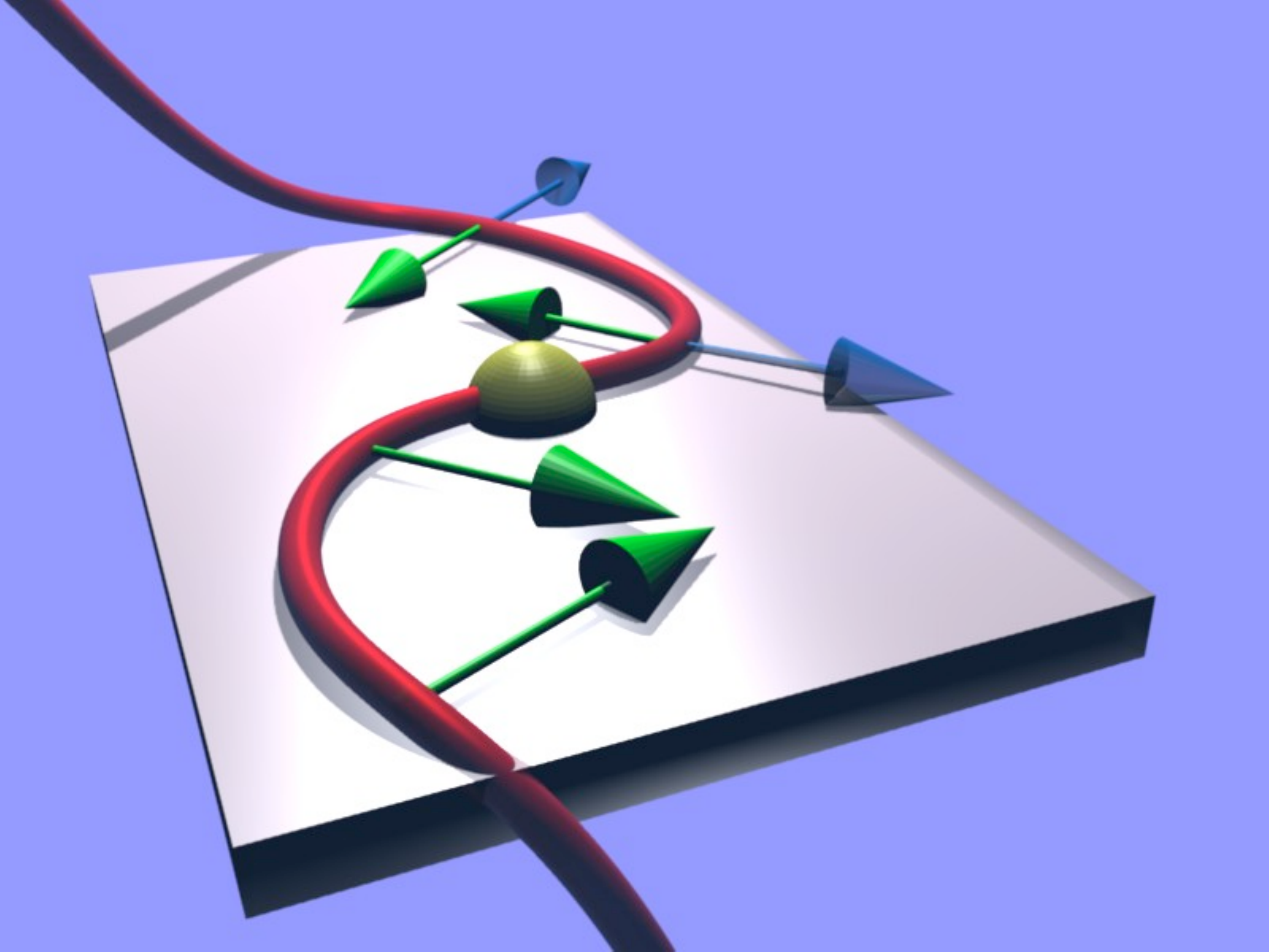}
        \caption{
      {A curve with inflection point (yellow ball). At each point the  direction of the (Frenet frame) normal vectors (green) is towards  the center of an oscullating circle.       
      There is a discontinuity in the direction of the normal vectors when we traverse the inflection point. At  this point the radius of the oscullating circle
      diverges 
      and the normal vector $\bf n$ becomes abruptly reflected in the oscullating plane from one side to the other side of the curve. The blue vector  
      equals opposite of the (reflected) normal vector $\bf n$ (see also Figure 6). }
                }
       \label{Figure 1:}
\end{figure}
Consequently a single non-degenerate inflection point  can not be 
removed by any local continuous deformation of the curve.  An isolated non-degenerate inflection point can only be locally and continuously removed
in the presence of another inflection point, by deforming the curve so that the inflection points  annihilate each other in a saddle-node bifurcation. 
In particular a sole non-degenerate inflection point 
can be removed only by translating it  away through an endpoint of the curve which involves a {\it global} deformation of the curve. 
This kind of stability enjoyed by an isolated inflection point under local deformations of the curve is the hallmark of a topological soliton.
Indeed,  let us recall the topological kink-soliton in a quartic double-well potential \cite{soliton}
\[
\ddot y \ = \ - \frac{d}{ds}  V(s) \  = \ - \frac{d}{ds} \left[ \frac{m^2}{2c^2} (y^2 - c^2)^2 \right]   \ = \   - \frac{2m^2}{c^2} y (y ^2 - c^2)
\]
\begin{equation}
y(s) \ = \  c \cdot \tanh[  m  ( s-s_0) ]
\la{ys}
\end{equation}
It describes a trajectory that interpolates between the two minima  $y = \pm c$ of the potential $V(s)$; See Figure 2. 
\begin{figure}[h]
        \centering
                \includegraphics[width=0.6\textwidth]{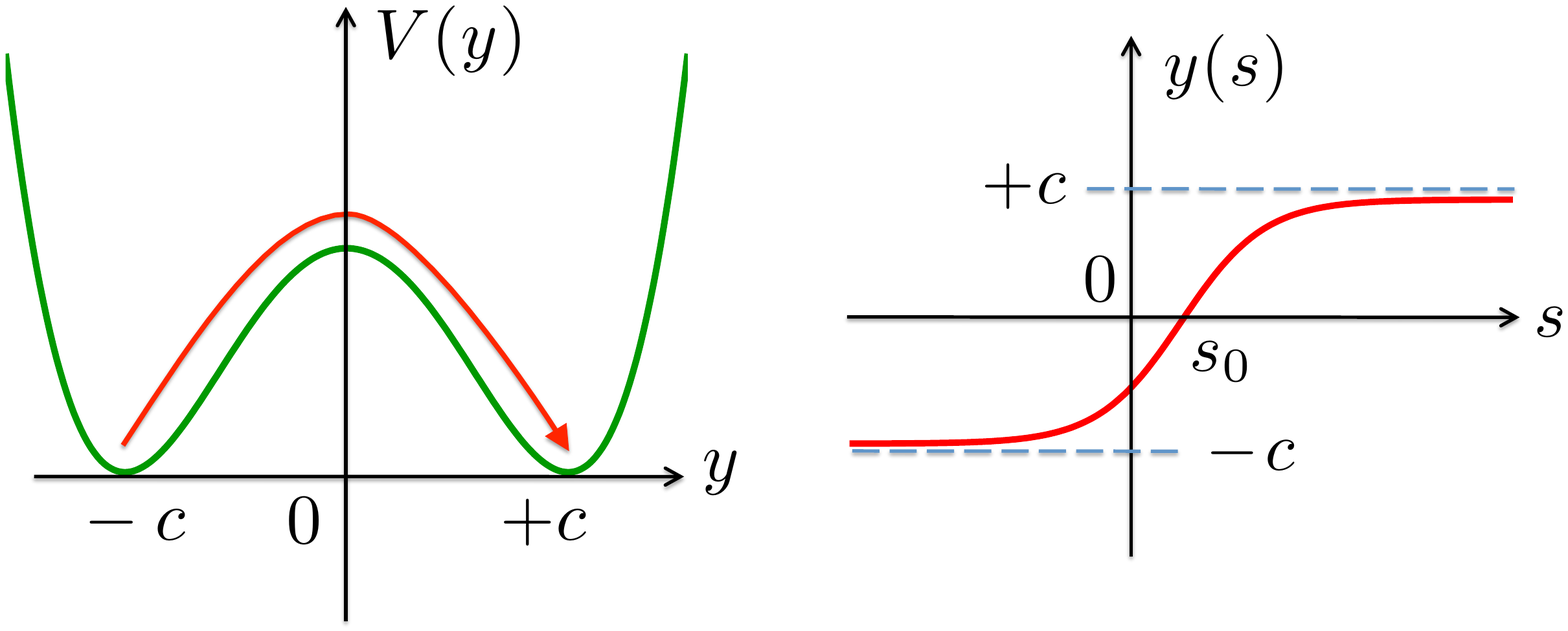}
        \caption{
      {The kink-soliton (right) interpolates between the two ground states at $\phi = \pm c$  of the potential  (left) as $s\to \pm \infty$. 
      It is topologically stable and can not be  removed by any finite energy deformation. }
                }
       \label{Figure 2:}
\end{figure}
The center of the
soliton is  at the point $s=s_0$ where $y(s)$ vanishes. The influence of this center point to the global topology of the trajectory can not be removed by 
any kind of continuous local deformation $y(s) \to y(s) + \delta y(s)$, as the resulting
curve continues to retain its characteristic global  property  that $y \to \pm c$ as $s\to \pm \infty$. Thus the deformed $y(s)$ 
necessarily vanishes at least at one point.
The goal of the present paper is to explain how this signature behaviour of a topological soliton can be detected and described in the case of discrete
piecewise linear curves, and in particular those curves that relate to the framing of folded proteins.

\section{{\bf II:} The Generalized Frenet Frame and Inflection Points}

\subsection{ {\bf A:} The Generalized Frenet Frame}

We start by describing the continuum Frenet equation and its generalizations.
Let $\mathbf x(s)$ be a space curve in $\mathbb R^3$. Its unit tangent vector
\[
\mathbf t \ = \   \frac{1}{ || \dot {\mathbf x} || } \, \dot {\mathbf x} \ \equiv \ \frac{1}{ || \dot {\mathbf x} || } \frac{ d \hskip 0.2mm \mathbf x (s)} {ds} 
\]
(we assume that $ || \dot {\mathbf x} ||  \not= 0$) is subject to the Frenet equation \cite{hansonbook}, \cite{kuipers}
\begin{equation}
\frac{d}{ds}\left(
\begin{matrix} 
{\bf n} \\
{\bf b} \\
{\bf t} \end{matrix} \right) =  || \dot {\mathbf x} || \left( \begin{matrix}
0 & \tau & -\kappa  \\ -\tau & 0 & 0 \\  \kappa & 0 & 0 \end{matrix} \right) 
\left(
\begin{matrix} 
{\bf n} \\
{\bf b} \\
{\bf t} \end{matrix} \right) 
\la{contDS1}
\end{equation}
where
\[
\mathbf b \ = \ \frac{ \dot {\mathbf x} \times \ddot {\mathbf x} } { || 
\dot {\mathbf x} \times \ddot {\mathbf x} || }
\]
is the unit binormal vector and 
\[
\mathbf n = \mathbf b \times \mathbf t
\] 
is the unit normal vector of the curve, and  
\[
\kappa(s) \ = \ \frac{ || \dot {\mathbf x} \times \ddot {\mathbf x} || } { || \dot  {\mathbf x} ||^3 }
\]
is the {\it frame independent} curvature of $\mathbf x(s)$ and
\[
\tau(s) \ = \ \frac{ (\dot {\mathbf x} \times \ddot {\mathbf x}) \cdot {\dddot {\mathbf x} }} { || \dot {\mathbf x} \times \ddot {\mathbf x} ||^2 }
\]
is the torsion.
The three vectors $(\mathbf n, \mathbf b, \mathbf t)$ form the right-handed orthonormal Frenet frame  at each point of the curve.

In the following we shall assume with no loss of generality, that $s \in [0,L]$ measures  the proper length along a curve with total length $L$ in $\mathbb R^3$ so that
\begin{equation}
|| \dot {\mathbf x} || = 1
\label{prole}
\end{equation}

Consider a curve with an isolated non-degenerate inflection point (or more generally a straight segment)  such as the one depicted in Figure 1.  
At the inflection point $s=s_0$  the Frenet frame can not be introduced since  $\kappa (s_0)$ vanishes; in the proper length 
gauge
\[
\kappa(s_0) = || \ddot {\mathbf x} (s_0) || = 0
\]
Conventionally, see {\it e.g.} \cite{spivak}, in the presence of inflection points  the Frenet equation (\ref{contDS1}) is usually introduced only piecewise between the inflection points,
for those values of $s$ for which $\kappa(s)$ is nonvanishing. But there are also alternative approaches that
allow for a continuous passage of the frame  through the inflection point (more generally straight segments).
For this we view the Frenet frame as an example of a general frame, obtained by starting from the
observation that while the tangent vector $\mathbf t(s)$ for a given curve is  unique, instead of $\{ \mathbf n(s), \mathbf b(s)\}$ 
we may choose an arbitrary orthogonal basis $\{\mathbf e_1(s), \mathbf e_2(s)\}$
for the normal planes of the curve that are perpendicular to $\mathbf t(s)$, without deforming the curve.  This general frame is  related to the Frenet frame 
by a local $SO(2)$ frame rotation around the frame independent tangent vector 
$\mathbf t(s)$  (see Figure 3),
\begin{figure}[h]
        \centering
                \includegraphics[width=0.4\textwidth]{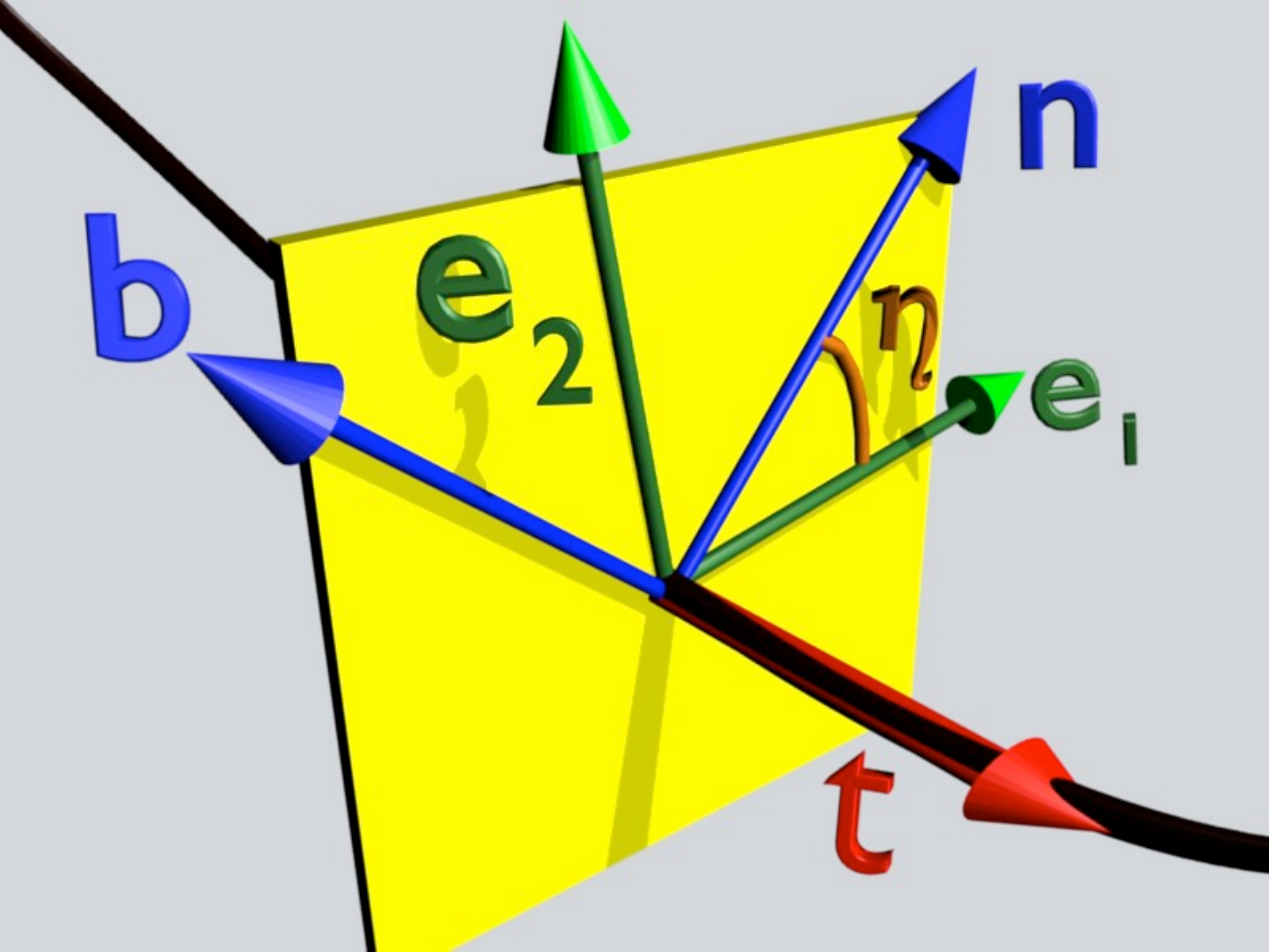}
        \caption{
      {The (blue) Frenet frame $(\mathbf n, \mathbf b)$ and a generic  (green) orthogonal frame $(\mathbf e_1 , \mathbf e_2)$ on the normal plane of $\mathbf t$,
      the tangent vector of the curve.}
                }
       \label{Figure 3:}
\end{figure}
\begin{equation}
\left( \begin{matrix} {\bf n} \\ {\bf b} \end{matrix} \right) \ \to \ \left( \begin{matrix} {{\bf e}_1} \\ {\bf e}_2 \end{matrix} \right) \
= \ \left( \begin{matrix} \cos \eta(s) & - \sin \eta(s) \\ \sin \eta(s) & \cos \eta(s) \end{matrix}\right)
\left( \begin{matrix} {\bf n} \\ {\bf b} \end{matrix} \right)
\la{newframe}
\end{equation}
The ensuing rotated version of the Frenet equation is
\begin{equation}
\frac{d}{ds} \left( \begin{matrix} {\bf e}_1 \\ {\bf e }_2 \\ {\bf t} \end{matrix}
\right) =
\left( \begin{matrix} 0 & (\tau - \dot \eta) & - \kappa \cos \eta \\ 
- (\tau - \dot \eta)  & 0 & -\kappa \sin \eta \\
\kappa \cos \eta & \kappa \sin \eta  & 0 \end{matrix} \right)  
\left( \begin{matrix} {\bf e}_1  \\ {\bf e }_2 \\ {\bf t} \end{matrix}
\right) 
\la{contso2}
\end{equation} 
If we recall  the adjoint basis of $SO(3)$ Lie-algebra
\begin{equation}
T^1 = \left( \begin{matrix} 0 & 0 & 0 \\
0 & 0 & -1 \\ 0 & 1 & 0 \end{matrix} \right) \ \ \ \ \  T^2 = \left( \begin{matrix} 0 & 0 & 1 \\
0 & 0 & 0 \\ -1 & 0 & 0 \end{matrix} \right) \ \ \ \ \ T^3 = \left( \begin{matrix} 0 & -1 & 0 \\
1 & 0 & 0 \\ 0 & 0 & 0 \end{matrix} \right)
\la{T}
\end{equation}
where
\[
[T^a , T^b ] = \epsilon^{abc} T^c
\]
we find that on  $\tau$ and $\kappa$ the  $SO(2)$ transformation acts as follows, 
\begin{equation}
\tau \ \to \ \tau - \dot \eta 
\la{sot}
\end{equation}
\begin{equation}
\kappa \ T^2  \ \to \ \kappa\  ( T^2 \cos \eta \ - T^1 \sin \eta ) \  \equiv \  e^{\eta T^3} ( \kappa T^2 )\,  e^{-\eta T^3}
\la{sok}
\end{equation}
If instead of $\eta \equiv 0$ that specifies the Frenet frame (Frenet gauge) we select $\eta(s)$ so that
\[
\eta(s) =  \int_0^s \! \tau (s') ds' 
\]
we arrive at Bishop's parallel transport frame \cite{bishop}; \cite{hansonbook}, \cite{kuipers} that
can be defined continuously and  unambiguously through inflection points.  
We note that  (\ref{sot}), (\ref{sok}) can be interpreted in terms of a $SO(2)$ gauge multiplet \cite{omaold}:  The change (\ref{sot}) in $\tau(s) $ is
identical to the $SO(2)\simeq U(1)$ gauge transformation of a one-dimensional gauge vector while $\kappa(s)$ transforms like  
a  component of a $SO(2)$ scalar doublet. This leads us to a gauge invariant quantity, the complex valued Hashimoto variable \cite{hasimoto} 
\begin{equation}
\xi(s) =  \kappa(s) \exp\left(i\int_0^{s} \! \tau \,ds'  \right) 
\la{hash1}
\end{equation}
When we combine (\ref{sot}) 
with a  $SO(2) \subset SO(3) $ rotation  (\ref{sok}) by $\eta(s)$ around 
the $T^3$-direction of the $SO(3)$ Lie algebra, the  effect on (\ref{hash1}) can  be summarized as follows,
\begin{equation}
\xi(s) \ \to \  \left[ \kappa(s) e^{-i \eta(s)} \right] \cdot \left[ \exp\left(i\int_0^{s} \! \tau \,ds'  + i \eta(s) \right) \right]  \cdot e^{ i \eta(0)}
\la{hash2}
\end{equation}
and thus the Hasimoto variable  $\xi(s)$  is manifestly independent of $\eta(s)$.  
(Note however, that the $\eta(0)$ dependence remains as an overall global 
phase ambiguity which is inherent to (\ref{hash2}) - the local gauge invariance becomes eliminated but a global one remains.)
In fact,  the Hasimoto variable simply combines the two 
real components of the $SO(2)$ scalar doublet
into a single complex valued  variable, with modulus that equals the frame independent {\it a.k.a.} gauge invariant geometric curvature of the curve.
In particular the Frenet frame is like the widely used  "Unitary Gauge" in the Abelian Higgs Model \cite{omaold}.

We find this language of gauge transformations in connection of frame rotations introduced in \cite{omaold}  
to be intuitively appealing and beneficial, and we shall use it frequently
in the sequel.

\subsection{ {\bf B:} Inflection Points}

We proceed to consider  a continuous curve  with  $n$  inflection points  at $s=s_i$,  
\[
s_0 = 0 < ... < s_i  < s_{i+1}  < ... < L= s_{n+1}
\]  
For simplicity we assume that the inflection points are isolated and non-degenerate zeroes of the curvature
\[
\kappa(s_i) \  = \ 0 
\]
A generalization to more involved inflection points is straightforward.
We take the curve to be of class $\mathcal C^3$. This ensures  that at each segment $(s_i, s_{i+1})$ the curvature is of class $\mathcal C^1$. Furthermore, 
since the inflection points are non-degenerate, as we approach an inflection point  the left and right derivatives of the curvature  
are non-vanishing and  in the limit when $s\to s_i$ they become equal in magnitude but have an opposite sign,  
\[
\frac{ d \kappa (s)}{ds}_{| s_i ^+} = - \frac{d  \kappa (s) }{ds}_{|  s_i^-}  \ \not= 0
\] 
This jump in the derivative of the curvature   is the signature of an inflection point in the Frenet frame.  
But  even though the curvature $\kappa(s)$ fails to be continuously differentiable the signed curvature
\begin{equation}
\tilde \kappa(s) \ = \  \sum_{i=0}^n (-1)^i \kappa(s) \theta(s-s_i)\theta(s_{i+1}-s)
\la{tilde}
\end{equation}
with $\theta(s)$  the unit step-function
\[
\theta(s) \  =  \  \left\{   \begin{matrix} 1 \   \  s>0 \\  0   \ \ s <0   \end{matrix}  \right.
\]
is now continuously differentiable  for all $s\in [0,L]$ and  and in particular
\[
\frac{ d \tilde \kappa }{ds}_{| s_i}  \not= 0 \ \ \ \ 
\]

The  original Frenet curvature $\kappa(s)$ and  the signed curvature $\tilde \kappa(s)$  are related by a gauge transformation (\ref{sok}) of
the Frenet frame, with  $\eta(s)$ given by the following  gauge transformation (\ref{sot}) of the Frenet torsion
\begin{equation}
\tau(s)  \ \to \ \tau(s)  - \dot \eta(s) \ = \ \tau(s) - \pi \cdot \frac{d} {ds} \sum_{i=1}^{n-1}  \theta(s-s_i) \ = \ \tau(s) - \pi  \sum_{i=1}^{n-1} \delta(s-s_i)
\la{contgau}
\end{equation}
This can be immediately verified by comparing the form of (\ref{tilde}) with that of the Hashimoto variable (\ref{hash1}), (\ref{hash2}).
We may call this gauge transformed version of the Frenet frame  the $\mathbb Z_2$-Frenet frame,   its discrete version will become important to us when 
we consider applications to folded proteins.

For a concrete example we take the plane curve  in Figure 1. For this curve, in the vicinity of the inflection point 
the Frenet curvature has clearly a qualitative form 
that may be described by the absolute value of the kink-soliton profile (\ref{ys}),
\[
\kappa(s) \ \sim \ \kappa_{0} \left| \tanh[m (s-s_0)]\right|
\]
Obviously the derivative of this curvature is discontinuous with a finite jump 
at the inflection point $\mathbf{\it I}$ where  $s=s_0$. This discontinuity reflects itself
in the abrupt change in the direction of the (green) normal vector $\bf n$, as depicted in Figure 1. 
The ensuing  {\it signed} curvature (\ref{tilde}) is qualitatively  described by the kink-soliton (\ref{ys})
\begin{equation}
\tilde \kappa(s) \sim  \ \kappa_{0} \tanh[m (s-s_0)]
\la{ys2}
\end{equation}
and  it  is manifestly continuously  differentiable, including the point $s=s_0$. Now the direction of the corresponding normal vector is also continuous 
through the inflection point. This is because the change in its direction
becomes compensated by the change in the sign of the signed  curvature when we cross the inflection point;  see the blue vectors in Figure 1, and Figure 6.

\section{{\bf III:} the discrete frenet equation}

\subsection{{\bf A:} The Discrete Frenet Frame}

In the sequel we  are primarily interested in  an open and oriented, piecewise linear discrete curve that we describe by a three-vector
${\bf r}(s) \in \mathbb R^3$.  The parameter
$s \in [0,L]$ measures the arc length and  $L$ is the total length of the curve. The curve
is determined by its vertices $C_i$ that are located at the positions
${\bf r}_i = ({\bf r}_0, \dots , {\bf r}_n)$   with ${\bf r}(s_i) = {\bf r}_i$. The endpoints of the curve are at
${\bf r}(0) = {\bf r}_0$ and 
${\bf r}(L) = {\bf r}_n$. The  nearest neighbor vertices $C_{i}$ and $C_{i+1}$ are connected by the line segments
\[
{\bf r}(s) \ = \ \frac{ s-s_{i} } {s_{i+1} - s_i} \, {\bf r}_{i+1}  \ - \ \frac{ s - s_{i+1}}  {s_{i+1} - s_i} \, {\bf r}_{i} 
\]
where $  s_{i} < s < s_{i+1}$. We 
utilize the Galilean invariance to translate the base of the curve 
to the origin in $\mathbb R^3$ so that
\[
{\bf r}_0 \ = \ 0
\]
The remaining global rotational orientation of the curve can then be fully determined by the choice of $\mathbf r_1$ and $ \mathbf r_2$.

For each pair of  nearest neighbor vertices  ${\bf r}_{i+1}$ and ${\bf r}_i$ along the curve we introduce the unit tangent vector 
\begin{equation}
{\bf t}_i = \frac{ {\bf r}_{i+1} - {\bf r}_i }{ | {\bf r}_{i+1} - {\bf r}_i |}
\la{t}
\end{equation} 
If all tangent vectors are known, the position of the $k^{th}$ vertex is  given by
\begin{equation}
{\bf r}_k \  = \ \sum_{i=0}^{k-1} |{\bf r}_{i+1} - {\bf r}_i| \cdot {\bf t}_i 
\la{DFE}
\end{equation}

We now introduce the discrete Frenet frame (DF frame) at the vertex $C_i$ at $\mathbf r_i$. This can be done 
whenever the three vertices at ${\bf r}_{i+1}$, ${\bf r}_i$ and ${\bf r}_{i-1}$ are not located on a common 
line so that ${\bf t}_i$ and ${\bf t}_{i-1}$ are not parallel. This enables us to determine 
the unit binormal vector
\begin{equation}
\hskip 3.0cm {\bf b}_i = \frac{ {\bf t}_{i-1} \times {\bf t}_i }{| {\bf t}_{i-1} \times {\bf t}_i|} \ \ \ \ \ (i = 1, ... , n-1)
\la{b}
\end{equation}
and the unit normal vector
\begin{equation}
{\bf n}_i = {\bf b}_i \times {\bf t}_i
\la{n}
\end{equation}
The orthogonal triplet $(\bf n_i , \bf b_i , \bf t_i)$ constitutes the discrete  
Frenet frame (DF frame) for the curve at the position of the vertex $\bf r_i$ for each $i=(1,...,n-1)$,
see Figure 4. 
\begin{figure}[h]
        \centering
                \includegraphics[width=0.4\textwidth]{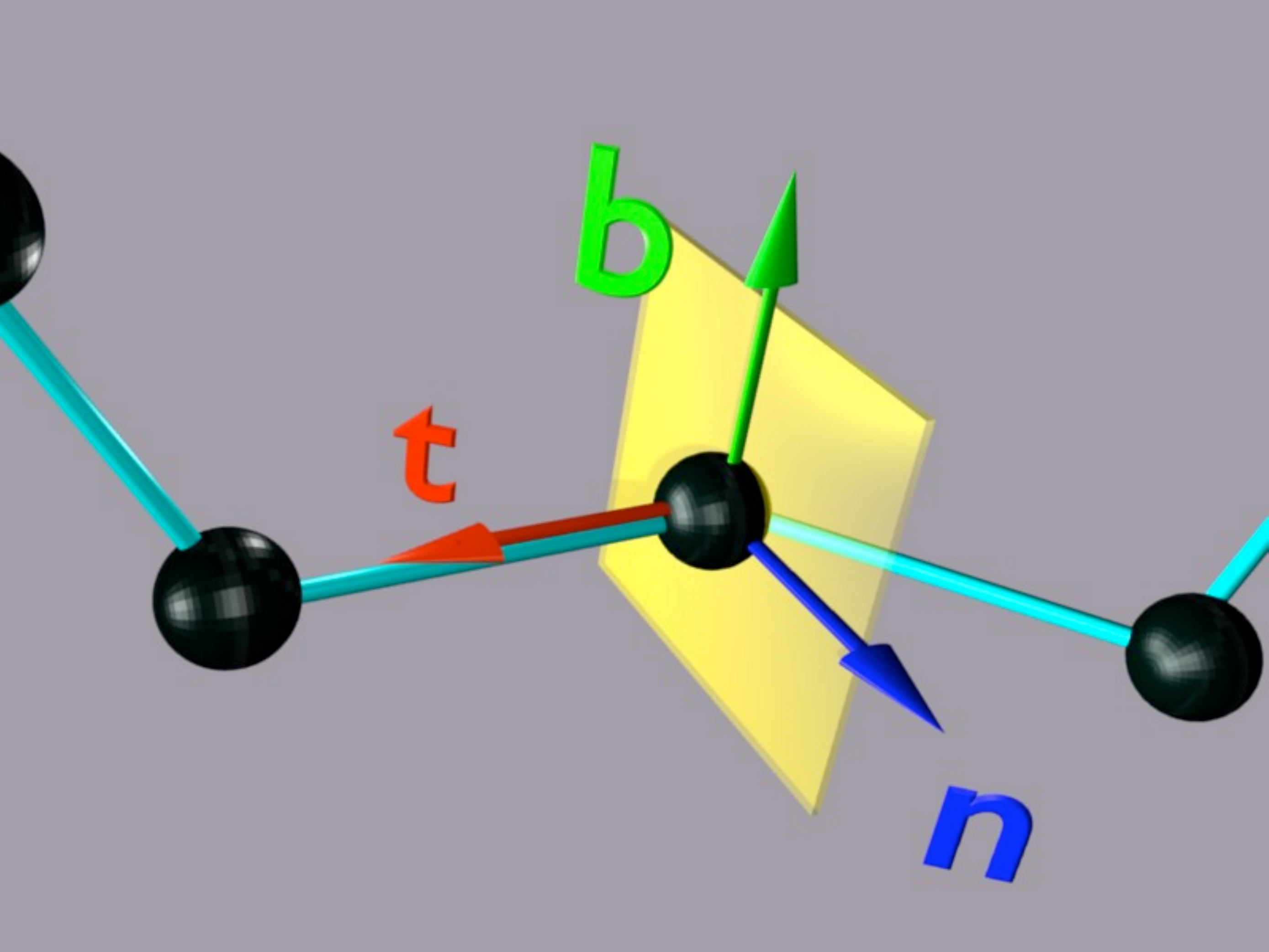}
        \caption{
      {A discrete piecewise linear curve is defined by its vertices $C_i$ and at each vertex there is an orthonormal discrete Frenet 
      frame (${\bf t}_i, {\bf n}_i, {\bf b}_i$), provided $\mathbf t_{i-1}$ and $\mathbf t_i$ are not parallel.}
                }
       \label{Figure 4:}
\end{figure}

\subsection{{\bf B:} The Transfer Matrix}

We now proceed  to derive a discretized version of the Frenet equation (DF equation) 
that relates the discrete Frenet frame at vertex $C_i$ to the discrete Frenet frame at vertex $C_{i+1}$ and
allows  for the construction of the curve  in terms of the
appropriate discrete versions of the curvature $\kappa(s)$ and torsion $\tau(s)$. 

From general considerations \cite{lattice} we conclude that the DF equation should
involve a transfer 
matrix $\mathcal R_{i+1,i}  $  that maps the  DF frame at the vertex $i$ to the DF frame at the vertex $i+1$,
\begin{equation}
\left( \begin{matrix} {\bf n}_{i+1} \\  {\bf b }_{i+1} \\ {\bf t}_{i+1} \end{matrix} \right)
\ = \ {\mathcal R}_{i+1,i}  \left( \begin{matrix} {\bf n}_{i} \\  {\bf b }_{i} \\ {\bf t}_{i} \end{matrix} \right)
\la{R}
\end{equation}
The construction of this transfer matrix then amounts to a solution of the DF equation: 
\[
\left( \begin{matrix} {\bf n}_{n} \\  {\bf b }_{n} \\ {\bf t}_{n} \end{matrix} \right) \ = \  
{\mathcal R}_{n,n-1} \cdot  {\mathcal R}_{n-1,n-2} \cdot  ... \cdot
{\mathcal R}_{2,1} \left( \begin{matrix} {\bf n}_{1} \\  {\bf b }_{1} \\ {\bf t}_{1} \end{matrix} \right)
\]
so that once the transfer matrix  is known for all  $i=1, ..., n-1$, we can use (\ref{R})
to construct all the Frenet frames for $i=2,....,n$ and the entire curve ${\bf r}(s)$ using (\ref{DFE}) 
together with the fact that the curve is linear in the intervals $s_{i-1} < s < s_i$;  We recall that for 
the initial conditions we need to specify  ${\bf r}_0$ that we have  already 
chosen to coincide with the origin ${\bf r}_0 = 0$, and  ${\bf r}_1$ and ${\bf r}_2$ that remove the degeneracy 
under global $SO(3)$ rotations of the curve in $\mathbb R^3$.

The transfer matrix  $ {\mathcal R}_{i+1,i}  $
is an element of the adjoint representation of $SO(3)$, thus we can parametrize it in terms of Euler angles. We choose the
($zxz$) angles 
\begin{equation}
 {\mathcal R}_{i+1,i} 
=  \left( \begin{matrix}
 -\sin\psi\sin\phi + \cos\theta \cos\psi\cos\phi & \sin\theta\cos\psi & 
-\sin\psi \cos\phi - \cos\theta\cos\psi\sin\phi \\
-\sin\theta \cos\phi & \cos\theta & \sin\theta \sin\phi \\
\cos \psi \sin \phi + \cos\theta \sin\psi \cos\phi & \sin\theta \sin\psi &
 \cos\psi \cos \phi - \cos\theta \sin\psi \sin \phi  \\
 \end{matrix} \right)_{\hskip -0.1cm i+1 , i} 
\la{expR}
\end{equation}
Here the angular variables have the following ranges: For the inclination angle $\theta$ 
we take  $\theta \in [0,\pi]\ {\rm mod}(2\pi)$  and for the two azimuthal angles we choose
$\phi \in [-\pi, \pi] \ {\rm mod}(2\pi)$ and $\psi  \in [-\pi, \pi] \ {\rm mod}(2\pi)$.
Note that since the angular variables are elements of the transfer matrix  that  takes the discrete Frenet frame from the vertex $i$ to the vertex
$i+1$, they are all to be interpreted as link variables that are defined on the bonds connecting the vertices. 

From (\ref{b}) we get  the following condition
\[
{\bf b}_{i+1} \cdot {\bf t}_i = 0
\]
Thus for each bond $(i,i+1)$
\[
\sin \theta \sin \phi \ = \ 0
\]
and we conclude from (\ref{t})-(\ref{n}) that for all $i$ we must have 
\[
\phi_{i+1,i} = 0
\]
This simplifies  the discrete Frenet equation into
\begin{equation}
\left( \begin{matrix} {\bf n}_{i+1} \\  {\bf b }_{i+1} \\ {\bf t}_{i+1} \end{matrix} \right)
\ = \
\left( \begin{matrix} \cos\psi \cos \theta & \cos\psi \sin\theta & -\sin\psi \\
-\sin\theta & \cos\theta & 0 \\
\sin\psi \cos\theta & \sin\psi \sin\theta & \cos\psi \end{matrix}\right)_{\hskip -0.1cm i+1 , i}
\left( \begin{matrix} {\bf n}_{i} \\  {\bf b }_{i} \\ {\bf t}_{i} \end{matrix} \right) \ \equiv \ {\mathcal R}_{i+1,i} 
\left( \begin{matrix} {\bf n}_{i} \\  {\bf b }_{i} \\ {\bf t}_{i} \end{matrix} \right)
\la{DFE2}
\end{equation}
Here
\begin{equation}
\cos\psi_{i+1 , i} = {\bf t}_{i+1} \cdot {\bf t}_i
\la{bond}
\end{equation}
is the discrete {\it  bond angle} and
\begin{equation}
\cos\theta_{i+1,i} = {\bf b}_{i+1} \cdot {\bf b}_i
\la{tors}
\end{equation}
is the discrete {\it torsion angle}. Geometrically, the bond angle $\psi_{i+1,i}$
measures the angle between ${\bf t}_{i+1}$ and ${\bf t}_i$ around ${\bf b}_{i+1}$
on the plane  that is determined by the three vertices $(C_i, C_{i+1}, C_{i+2})$ (Figure 5). 
The torsion angle $\theta_{i+1, i}$ measures the angle between the two planes that are determined by the 
vertices $(C_{i-1}, C_{i}, C_{i+1})$ and $(C_i, C_{i+1}, C_{i+2})$, respectively (Figure 5).
\begin{figure}[h]
        \centering
                \includegraphics[width=0.4\textwidth]{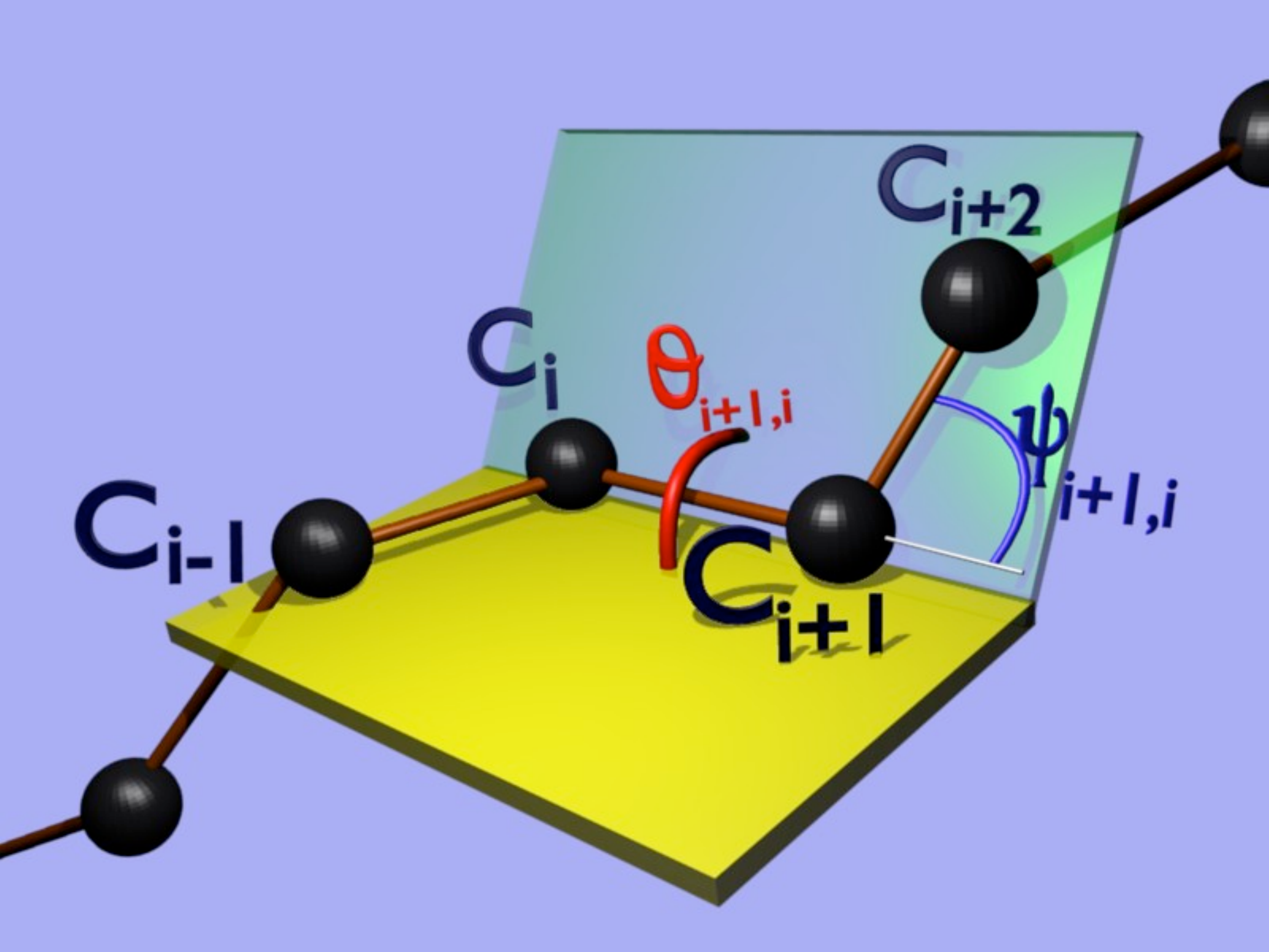}
        \caption{
      {The bond angle $\psi_{i+1,i}$ is determined by the three vertices $(C_{i-1}, C_{i}, C_{i+1})$. The torsion angle $\theta_{i+1, i}$ 
      is the angle between the two planes determined by
vertices $(C_{i-1}, C_{i}, C_{i+1})$ and $(C_i, C_{i+1}, C_{i+2})$}
                }
       \label{Figure 5:}
\end{figure}
We give these planes an orientation in $\mathbb R^3$ 
by extending the range of the torsion angle from $\theta_{i+1,i} \in [0,\pi]$ into $\theta_{i+1,i} \in [-\pi, \pi] \  {\rm mod}(2\pi)$. 
This introduces a discrete $\mathbb Z_2$ symmetry 
\begin{equation}
\mathbb Z_2 \ : \ \  \theta_{i+1,i} \ \leftrightarrow \ -  \theta_{i+1,i}
\la{Z2}
\end{equation}
that we find useful in the sequel.

We recall  the Rodrigues formula
\begin{equation}
e^{\alpha \mathbb U} = \mathbb I + \mathbb U \sin \alpha + \mathbb U^2 ( 1-\cos \alpha)
\end{equation}
where 
\[
\mathbb U \ = \ {\bf u} \cdot {\bf T} \ = \ u^aT^a
\]
and $T^a$ are the $SO(3)$ matrices (\ref{T}) and
$\bf u$ is a unit vector.
With these we can write the transfer matrix as follows,
\begin{equation}
{\mathcal R}_{i+1,i}  \ = \
\exp\{ - \psi_{i+1,i} T^2 \} \cdot \exp \{ - \theta_{i+1,i} T^3 \} 
\ = \ \exp\{-\alpha {\bf v} \cdot {\bf T}\}_{i+1,i} 
\la{trans}
\end{equation}
where 
\[
\alpha_{i+1,i} = 2 \arccos \left[ \frac{1}{4} ( {\bf b}_{i+1} \cdot {\bf b}_i)
( {\bf t}_{i+1} \cdot {\bf t}_i ) \right]
\]
and 
\[
{\bf v}_{i+1,i} = \frac{1}{\sin \frac{\alpha}{2}}
\left( - \sin \frac{\psi}{2} \sin {\frac{\theta}{2}} \ ,  \ \sin \frac{\psi}{2} \cos \frac{\theta}{2} \  , \ 
\cos \frac{\psi}{2} \sin \frac{\theta}{2} \right)_{i+1,i}
\]

\subsection{{\bf C:} Gauge symmetries}

Let us consider the effect of the discrete version of the
local $SO(2)$ rotation (\ref{newframe}),
\begin{equation}
 \left( \begin{matrix}
{\bf n} \\ {\bf b} \\ {\bf t} \end{matrix} \right)_{i}
\rightarrow  e^{\Delta_i T^3} \left( \begin{matrix}
{\bf n} \\ {\bf b} \\ {\bf t} \end{matrix} \right)_{i}
\la{discso2}
\end{equation}
For the covariance of the DF equation under  (\ref{discso2}) we need 
\begin{equation}
e^{-\theta_{i+1,i} T^3} \to e^{\Delta_{i+1} T^3 } \cdot e^{-\theta_{i+1,i} T^3} e^{-\Delta_i T^3}
\end{equation}
\begin{equation}
e^{-\psi_{i+1,i} T^2 } \to e^{\Delta_{i+1} T^3} \cdot e^{-\psi_{i+1,i} T^2} \cdot e^{- \Delta_{i+1}T^3}
\end{equation}
A direct computation shows that this implies the following transformation laws
\begin{equation}
\theta_{i+1,i} \to \theta_{i+1,i} + \Delta_i - \Delta_{i+1}
\la{gaugetheta}
\end{equation}
\begin{equation}
\psi_{i+1,i} T^2 \to \psi_{i+1,i} ( T^2 \cos \Delta_{i+1} - T^1 \sin \Delta_{i+1} )
\la{gaugepsi}
\end{equation}
These are the discrete versions of the transformations of $\tau$ and $\kappa$  in (\ref{sot}), (\ref{sok})
respectively.  

Explicitely, the gauge transformed transfer matrix is
\begin{equation}
e^{\Delta_{i+1} T^3} \!  {\mathcal R}_{i+1,i} \,  e^{-\Delta_i T^3} \! \!  \! \equiv  
 {\mathcal R}^\Delta_{i+1,i}
 \la{RD}
 \end{equation}
\begin{equation}
= \left( \begin{matrix} \cos \Delta  \cos \theta_\Delta   \cos \psi + \sin\Delta  \sin \theta_\Delta  \hskip 0.1cm &
\cos \Delta \sin \theta_\Delta   \cos \psi - \sin \Delta \cos \theta_\Delta  \hskip 0.1cm  & - \cos \Delta \sin \psi 
\\
\sin \Delta  \cos \theta_\Delta  \cos \psi -  \cos\Delta \sin \theta_\Delta  \hskip 0.1cm &
\sin \Delta  \sin \theta_\Delta \cos \psi + \cos \Delta \, \cos \theta_\Delta \hskip 0.1cm  & - \sin \Delta\sin \psi
 \\
\cos \theta_\Delta \, \sin\psi & 
\sin\theta_\Delta \, \sin\psi & \cos \psi 
\\
 \end{matrix} \right)_{\hskip -0.2cm i+1 , i} 
\la{expRd}
\end{equation}
We have here  used the notation
\begin{equation}
\begin{matrix}
\Delta \ &  \equiv \ & \Delta_{i+1} \\
\theta_\Delta \  & \equiv \ &  \theta_{i+1,i} + \Delta_i
\end{matrix}
\la{thetaD}
\end{equation}
and the corresponding general frame Frenet equation is
\begin{equation}
\left( 
\begin{matrix} 
{\bf e}_1  \\  {\bf e }_2 \\ {\bf t}
\end{matrix} \right)_{i+1}
= \  {\mathcal R}^\Delta_{i+1,i} 
\left( \begin{matrix} {\bf e}_{1} \\  {\bf e }_{2} \\ {\bf t} \end{matrix} \right)_i
\la{DFE3}
\end{equation}
Notice that even though the explicit matrix elements in (\ref{expRd}) do not have a manifestly covariant form in terms of the link variables, 
the gauge transformed transfer matrix (\ref{RD}) is by construction a  covariant  link variable.

\subsection{{\bf D:} Continuum Limit }

The different choices of $\Delta_i$ in (\ref{DFE3}) correspond to different generalized Frenet frames.
We shall now verify that with the general version of  transfer matrix (\ref{expRd}), this indeed yields the generalized Frenet equation (\ref{contso2}) in
the continuum limit where the distances between the vertices $C_i$ of the curve vanish, provided the limit is a class $\mathcal C^3$ curve.
\[
| {\bf r}_{i+1} - {\bf r}_i| \ \approx  \ \epsilon \ \to \ 0
\]
We define
\begin{equation}
\begin{matrix}
\psi_{i+1,i}  & =  \ \epsilon  \cdot \kappa_{i+1,i}  \\  
\theta_{i+1,i}  & = \ \epsilon \cdot \tau_{i+1,i} \\
\Delta_{i+1} -  \Delta_{i} &  = \  \epsilon \cdot \sigma_{i+1, i}  \\
\hskip .2cm \frac{1}{2} (\Delta_{i+1} +  \Delta_{i}) &  \hskip -0.35cm  = \ \eta_{i+1,i} 
\end{matrix}
\la{ident}
\end{equation}
where $\sigma_{i+i,i}$ are some finite constants.  When we expand  (\ref{DFE3}) in $\epsilon$ we get in the leading order
\begin{equation}
\frac{1}{\epsilon} \left[ \ \left( \begin{matrix} {\bf e}_1  \\  {\bf e }_2 \\ {\bf t}  \end{matrix} \right)_{i+1}\! \!  \! - \ 
\left( \begin{matrix} {\bf e}_1 \\  {\bf e }_2 \\ {\bf t}  \end{matrix} \right)_i \ \right]
= \left( \begin{matrix} 0 & ( \tau - \sigma ) & -\kappa \cos \eta \\
- (\tau - \sigma ) & 0 & -\kappa \sin \eta  \\
\kappa \cos \eta & \kappa \sin \eta  & 0 \end{matrix} \right)_{\hskip -0.1cm i+1,i} \left( \begin{matrix} {\bf e}_{1} 
\\  {\bf e }_2 \\ {\bf t}  \end{matrix}\right)_i 
\la{contfre}
\end{equation}
If the $\epsilon \to 0$ exists it gives us the 
generalized continuum Frenet equation (\ref{contso2}), with the identification 
\[
\sigma \ \to \ \dot \eta
\]
and  the identification (\ref{ident}) between the discrete torsion and curvature angles with their continuum counterparts.

\subsection{ {\bf E:} Inflection points}

Consider a piecewise linear curve that has a single isolated inflection point located at vertex $C_i$; A generalization to several inflection points and
straight segments
is straightforward. By assumption, the preceding vertex $C_{i-1}$ admits a Frenet frame.  Since the tangent vectors  $\mathbf t_i$ and $\mathbf t_{i-1}$ are parallel, 
at  the vertex $C_i$ both the normal vector  $\mathbf n_i$  and the binormal vector $\mathbf b_i$  of a Frenet frame can not be determined and the
Frenet frame at $C_i$ can not be introduced. Consequently the torsion angle $\theta_{i,i-1}$ can not be defined. But the definition of
the bond angle involves only the tangent vectors so it can still be computed and from (\ref{bond}) we get 
\[
\psi_{i,i-1} = 0 \ \ \  ({\rm mod }\ 2\pi)  
\]
In order to introduce a framing of the curve that covers the  vertex $C_i$, we proceed as follows: We first deform the curve 
slightly by moving the vertex $C_i$ in a direction of some  arbitrarily chosen vector ${\mathbf u}$ that is not parallel with $\mathbf t_i$,
\begin{equation}
\mathbf r_i \ \to \  \mathbf r_i + \epsilon \cdot \mathbf u
\la{shiframe}
\end{equation}
Here the limit $\epsilon \to 0$ is tacitly understood. The introduction of  $\mathbf u$  removes the inflection point from  the shifted vertex $\tilde C_i$ and
this enables us to introduce a $\mathbf u$ dependent Frenet frame at the shifted vertex $\tilde C_i$. In the limit where $\epsilon$ vanishes
we get a $\mathbf u$ dependent  frame  at the original vertex $C_i$, obtained  by transferring the Frenet frame from the vertex  $C_{i-1}$ as follows,
\begin{equation}
\left( \begin{matrix} {\bf e}_1  \\  {\bf e }_2 \\ {\bf t}  \end{matrix} \right)_{i} 
= \left( \begin{matrix} \cos \hat \theta  & \sin \hat \theta   \hskip 0.2cm & 0  \ \\
- \sin \hat \theta   \hskip 0.35cm & \cos \hat \theta  & 0 \\
0 & 0 & 1  \end{matrix} \right)_{\hskip -0.1cm i,i-1}\! \!  \left( \begin{matrix} {\bf n} 
\\  {\bf b } \\ {\bf t}  \end{matrix}\right)_{i-1} 
\la{gtrs}
\end{equation}
Here $\hat \theta_{i,i-1}$ is now some description {\it i.e.} explicitely $\mathbf u$ dependent angle. 

In order to establish that the frame can be chosen in
a $\mathbf u$ independent manner we proceed to 
remove the explicit $\mathbf u$ dependence.  For this we introduce the gauge transformation (\ref{gaugetheta}) in (\ref{gtrs}) which sends
\[
\hat \theta_{i,i-1} \ \to \ \hat \theta_{i,i-1} + \Delta_{i-1} - \Delta_{i}
\]
Since  we have the original Frenet frame at the vertex $C_{i-1}$, we also have
\[
\Delta_{i-1} = 0
\] 
But $\Delta_i $ is freely at our disposal and we may choose it so that any $\mathbf u$ dependence becomes removed. This leaves
us with a $\mathbf u$ independent  reminder that we may choose at our convenience,
\[
\hat \theta_{i,i-1} - \Delta_{i}   \ \equiv \  \hat \Delta_{i,i-1}
\]
where $\hat \Delta_{i,i-1}$ is now by construction a $\mathbf u$ independent quantity, at our disposal. Different choices correspond to different
gauges.

Since $\mathbf t_i$ and $\mathbf t_{i+1}$ are not parallel,  we can proceed to construct a frame at vertex $C_{i+1}$ from the frame
$(\mathbf e_1, \mathbf e_2, \mathbf t)_i$ at vertex $C_i$ using the transfer matrix (\ref{expRd}). Since the remaining gauge parameters  $\Delta_k$
with $k>i$  are all at our disposal,  we may return to the Frenet frame, or select any other convenient framing, at the vertex $C_{i+1}$ and 
at all subsequent vertices. If the goal is to approximate a continuous space curve, in  the limit of vanishing bond length the gauge
parameters $\Delta_k$ should be selected in such a manner that in the continuum limit they yield the gauge function $\eta(s)$ and so that
the ensuing discrete transfer matrix  smoothly goes over to its
continuum limit (\ref{contfre})

\subsection{{\bf F:} Discrete gauge transformations}

The transfer matrix ${\mathcal R}_{i+1,i}$ determines the curve in $\mathbb R^3$ up to rigid Galilean motions {\it i.e.} global
translations and spatial rotations. 
The improper spatial rotation group $O(3)$ acts on each of the vertices ${\bf r}_k$ in (\ref{DFE}) by a rotation matrix $\mathcal O \in O(3)$ that sends each of the ${\bf r}_k$ into
\[
{\bf r}_k \ \to \ {\mathcal O} {\bf r}_k
\]
As a consequence only the global orientation of the curve in $\mathbb R^3$ changes.
An example  is the improper rotation that inverts the curve in $\mathbb R^3$ 
by reversing the direction of each tangent vector
\[
{\bf t}_i \ \to - {\bf t}_i
\]
but with no effect on the ${\bf n}_i$ and ${\bf b}_i$. From the explicit form of the transfer matrix in (\ref{DFE2}) we
conclude that this corresponds to  the following global version of (\ref{gaugetheta}), (\ref{gaugepsi})
\[
\begin{matrix} \theta_i  \\ \psi_i  \end{matrix}  \  \begin{matrix} \to \\ \to \end{matrix} \
\begin{matrix} \theta_i \\ - \psi_i \end{matrix}
\]
That is, $\Delta_i = \pi$ for all $i$. 
Consequently if we include this improper rotation in our gauge structure
we can restrict the range of $\psi_i$ from $\psi_i \in  [-\pi, \pi] \ {\rm mod}(2\pi)$ to $\psi_i \in [0,\pi] \ {\rm mod}(2\pi)$,
but we prefer to continue with the extended range. 

Similarly, we can introduce the improper rotation that sends
\[
{\bf b}_i \ \to  \  - {\bf b}_i
\]
with no effect on ${\bf t}_i$ and ${\bf n}_i$. Since the ${\bf t}_i$ remain intact the curve does not change, and from the DF equation
(\ref{DFE2}) we conclude that this corresponds to the following global $\mathbb Z_2$ transformation
\[
\begin{matrix} \theta_i  \\ \psi_i  \end{matrix}  \  \begin{matrix} \to \\ \to \end{matrix} \
\begin{matrix} - \theta_i \\ \psi_i \end{matrix}
\]
This is the $\mathbb Z_2$ symmetry that we have introduced in (\ref{Z2}), to extend the 
range of $\theta_i$ from $\theta_i \in  [0, \pi] $ to $\theta_i \in [-\pi,\pi] \ {\rm mod}(2\pi)$. We note that this symmetry of the underlying
curve can not be reproduced by the gauge transformation (\ref{gaugetheta}), (\ref{gaugepsi}), nevertheless the curve remains intact since
the $\mathbf t_i$ do not change.

Another useful discrete transformation in our subsequent discrete curve analysis is the proper rotation that at a given vertex $C_i$ sends 
\[ 
\begin{matrix}
{\bf b}_i \ \to  \  - {\bf b}_i \\ 
{\bf n}_i \ \to \ - {\bf n}_i 
\end{matrix}
\]
but with no effect on ${\mathbf t}_i$ so that the curve remains intact.  
This rotation is obtained by selecting $\Delta_{i+1} = \pi$ and with all  $\Delta_{k} = 0$ at the preceding
vertices $C_k$ (with $k \leq i$). Since the $\Delta_{i+1}$ 
appears in the gauge transformation law of both $\theta_{i+1,i}$ and $\theta_{i+2,i+1}$, this leads  to the following
realization of the gauge transformation (\ref{gaugetheta}), (\ref{gaugepsi})
 \[
\begin{matrix}
\theta_{i+1,i}  & \to &  \  \theta_{i+1,i} - \pi  \\
\theta_{i+2,i+1}  & \to & \  \theta_{i+1,i} + \pi  \\
\psi_{i+1,i} & \to  & \  - \psi_{i+1,i} 
\end{matrix}
\]
If we generalize this gauge transformation  by selecting 
\[
\Delta_{k} = \pi \ \ \ \ \ \ {\rm for } \  \ k\geq i+1
\]
with
\[
\Delta_{k} = 0 \ \ \ \ \ \ {\rm for } \  \ k < i+1
\]
where the vertex $C_i$ is preselected, the gauge transformation becomes
\begin{equation}
\begin{matrix}
\theta_{i+1,i \ }  & \to &  \hskip -2.5cm \theta_{i+1,i} - \pi  \\
\psi_{k+1,k} & \to  &  - \ \psi_{k+1,k} \ \ \ \hskip 1.0cm  {\rm for \ \ all} \ \  k \geq i 
\end{matrix}
\la{dsgau}
\end{equation}
Since the bond angle is the discrete version of the Frenet curvature (\ref{ident}), we recognize here the discrete
analog of the continuum gauge transformation (\ref{tilde}), (\ref{contgau}).  For a piecewise linear discretization of  a plane curve
such as the one Figure 1, this enables us to introduce a framing that captures the kink-soliton
behaviour  (\ref{ys}), (\ref{ys2}) of the inflection point, with the change of sign in curvature at the soliton position (Figure 6).
\begin{figure}[h]
        \centering
                \includegraphics[width=0.6\textwidth]{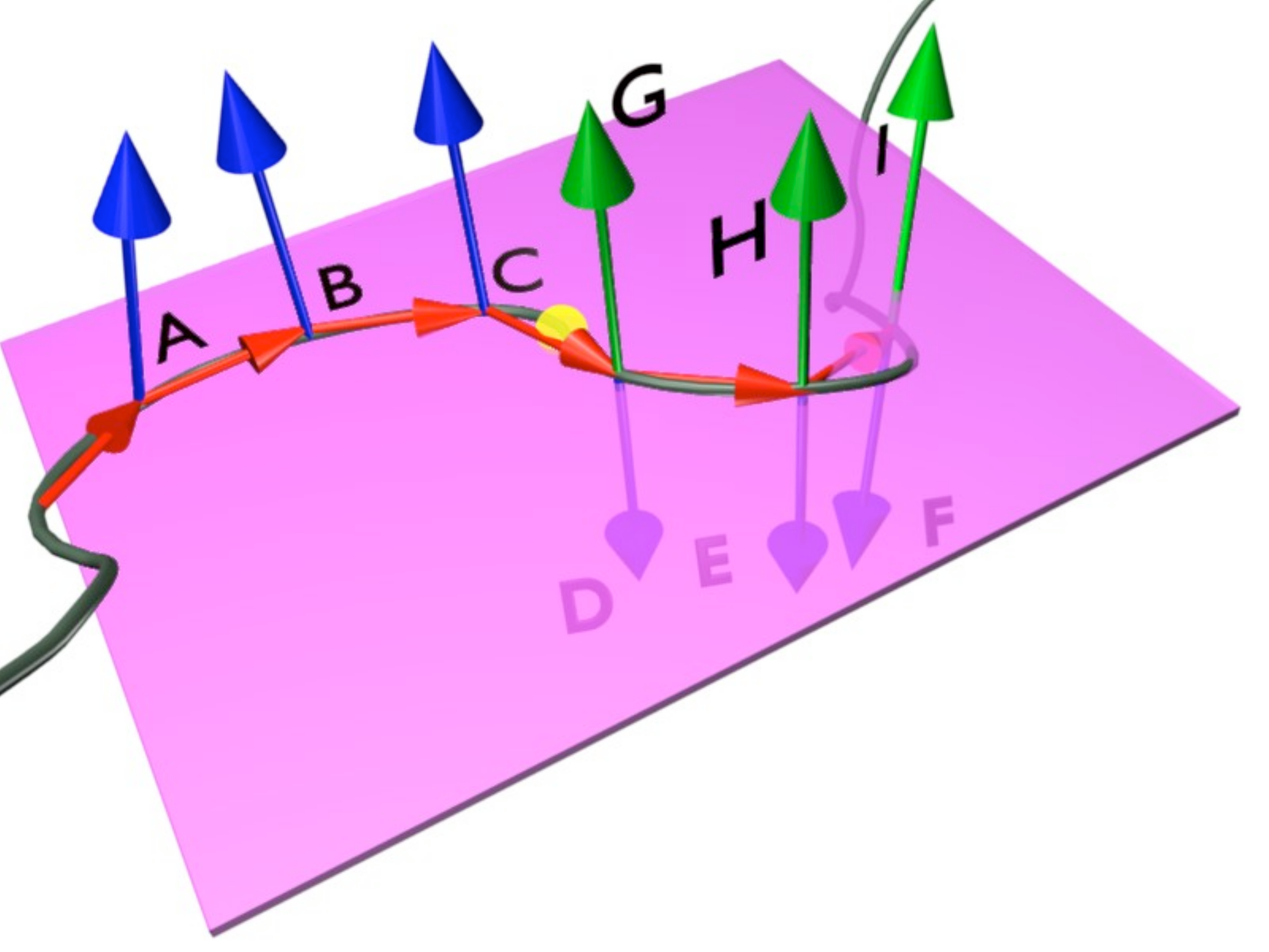}
        \caption{
      { A continuous plane curve with an inflection point (yellow dot) such as the one in Figure 1,  together  with its discrete approximation. 
      The tangent vectors $\mathbf t_i$ (red) of the discrete approximation can be chosen so that two neighbors are never parallel
      and thus a discrete Frenet frame can be introduced at each vertex. When we pass through the inflection point the direction of the
      binormal vectors following ($\bf A,B,C$) becomes reflected in the plane  into ($\bf D,E,F$) and  there is a discontinuity in the Frenet framing.
       But if we introduce the gauge transformation (\ref{dsgau}) at  vertices after the inflection point, the ensuing framing ($\bf A,B,C,G,H,I$) is 
       continuous.
      }
                }
       \label{Figure 6:}
\end{figure}

\subsection{G: Curve Construction}

An example of problems where the present formalism can be applied  is the construction of a discrete and piecewise linear curve from the
known values of its bond and torsion angles. These angles can be constructed  for example using an energy principle to locate
a minimum energy configuration of some energy functional 
\[
E(\psi_{k+1,k}, \theta_{k+1,k})
\]  
We may define the angles using the 
Frenet frame.  Examples of energy functionals have been discussed in \cite{omaold}, \cite{ulf}. 

Three vertices are needed to specify the position and the overall rotational orientation of the curve.
To compute  a single bond angle from the curve, we need three vertices while for the torsion angle we need four; See Figure 5.
Consequently from the first three initial positions of the curve, $(\mathbf r_{0}, \mathbf r_1, \mathbf r_2)$, we can compute the first bond angle $\psi_{1,0}$.
But in order to compute the first pair $(\psi_{2,1}, \theta_{2,1})$ we also need to specify $\mathbf r_3$.  

Here we are interested in the inverse problem where the set of angles $\{ \psi_{k+1,k}, \theta_{k+1,k}\} $ are assumed to be
known. Depending on the boundary conditions for the energy functional, the known initial data may also include numerical values of $(\psi_{1,0}, 
\theta_{1,0})$, even though $\theta_{1,0}$ lacks a geometric interpretation.  In such a  case we can immediately proceed 
to the computation of the entire curve using (\ref{DFE2}) or alternatively using the transfer matrix (\ref{DFE3}), starting from   
an initial choice of  frame $(\mathbf n_0, \mathbf b_0, \mathbf t_0)$. Different initial choices are related to each other by {\it global} {\it i.e.}
index $i$ independent parameter $\Delta$ in (\ref{gaugetheta}), (\ref{gaugepsi}).

We get both the frame at  the vertex $k$ and its location $\mathbf r_k$ 
when we also employ  (\ref{DFE}),  starting from a given initial value $\mathbf r_0 (=0)$. 

In general we expect to have a situation where the three first points  $(\mathbf r_{0}, \mathbf r_1, \mathbf r_2)$ are given.
From these points we get the two tangent vectors $\mathbf t_0$ and  $\mathbf t_1$. We then use (\ref{b}), (\ref{n}) to complete 
the  Frenet frame at the location $\mathbf r_1$.  We identify the bond angle $\psi_{1,0}$ with the angle between the 
two vectors $\mathbf t_0$ and $ \mathbf t_1$ using (\ref{bond}). This bond angle may or may not be
determined by the energy functional. If it is determined, the angle between  $\mathbf t_0$ and $ \mathbf t_1$ is determined and instead of 
fully specifying $\mathbf r_2$ we only need to specify its distance from $\mathbf r_1$ and the remaining directional angle  that we may call 
$\theta_{1,0}$.

For a  practical algorithmic implementation the following choice can be convenient,
\[
\mathbf r_0 =  \delta_{1,0}  \left( \begin{matrix} - \cos \psi_{1,0}  \\  \sin \psi_{1,0} \\  0 \end{matrix} \right)
\]
\[
\mathbf r_1 = \left( \begin{matrix}  0 \\ 0 \\ 0 \end{matrix} \right)
\]
\[
\mathbf t_0 = \left( \begin{matrix} \cos \psi_{1,0} \\  -\sin \psi_{1,0} \\ 0 \end{matrix} \right)
\]
\begin{equation}
\mathbf n_1 = \left( \begin{matrix} 1 \\ 0 \\ 0 \end{matrix} \right) \ \ \ \ \ \ 
\mathbf b_1 =   \left( \begin{matrix} 0 \\ 1 \\ 0 \end{matrix} \right)
\ \ \ \ \ \ 
\mathbf  t_1 = \left( \begin{matrix} 0 \\ 0 \\ 1 \end{matrix} \right)
\la{frenetini}
\end{equation}
where we have introduced the notation 
\[
\delta_{k+1,k}=  |\mathbf r_{k+1} - \mathbf r_{k}|
\]
for the segment lengths.
The generalized  Frenet frame together with the corresponding 
location of the vertex $\mathbf r_{i+1}$ can then be computed by iterative application
of 
\begin{equation}
\left( \begin{matrix} \mathbf  n \\
\mathbf b \\
\mathbf t \\
\mathbf r 
\end{matrix}
\right)_{i+1} \! \! \! \! \! =  \
\mathcal T_{i+1, i}
 \left( \begin{matrix} \mathbf n \\
\mathbf b \\
\mathbf t \\
\mathbf r 
\end{matrix}
\right)_{i}  
= \ 
\left( \begin{matrix}
&   &   & 0 \\
 & \Biggl(  {\huge{  \mathcal R}} \Biggr) &   & 0 \\
 &    &  & 0  \\ 
0 & 0& \delta & 1 \\
\end{matrix} \right)_{i+1,i}     
\left( \begin{matrix} \mathbf n \\
\mathbf b \\
\mathbf t \\
\mathbf r 
\end{matrix}
\right)_{i+1}
\label{geneT}
\end{equation}
This can be directly generalized into 
\[
\left( \begin{matrix} \mathbf  e_1 \\
\mathbf e_2  \\
\mathbf e_3 \\
\mathbf r 
\end{matrix}
\right)_{i+1} \! \! \! \! \! =  \
\mathcal T^\Delta_{i+1, i}
 \left( \begin{matrix} \mathbf e_1 \\
\mathbf e_2 \\
\mathbf e_3 \\
\mathbf r 
\end{matrix}
\right)_{i}  
= \ 
\left( \begin{matrix}
&   &   & 0 \\
 & \Biggl(  {\huge{  \mathcal R}^\Delta} \Biggr) &   & 0 \\
 &    &  & 0  \\ 
\delta_1 & \delta_2 & \delta_3 & 1 \\
\end{matrix} \right)_{i+1,i}     
\left( \begin{matrix} \mathbf e_1 \\
\mathbf e_2 \\
\mathbf e_3 \\
\mathbf r 
\end{matrix}
\right)_{i+1}
\]
where $\mathcal R^\Delta$ is the matrix (\ref{expRd})
and the $\delta_{1}, \delta_2, \delta_3 $ are the components of the vector
\[
{\vec {\mathbf \delta}}_{k+1,k}  = \delta_{k+1,k} \cdot \left( \begin{matrix} \cos \alpha \sin \beta \\ \sin \alpha \sin \beta \\ cos \beta \end{matrix} \right)_{k+1,k}
\]
When $\beta = 0$ (and $\Delta = 0$) we obtain the transfer matrix (\ref{geneT}) with  $\mathbf t_k$ the tangent vector of the curve, 
while for general $(\alpha,\beta)$ the tangent of the curve is in the direction of $\vec \delta$ in the $(\mathbf e_1, \mathbf e_2, \mathbf e_3)$ 
frame. Thus this transfer matrix provides a rule for
transporting an {\it a priori} arbitrarily oriented orthogonal frame along the curve.

Of particular interest is the construction of a discrete version of  Bishop's parallel transport frame \cite{bishop}, as a gauge transformed version of
the discrete Frenet frame. Since the Frenet frame  starts with $(\psi_{2,1}, \theta_{2,1})$ and  can be constructed once 
$(\mathbf r_{0}, \mathbf r_1, \mathbf r_2, \mathbf r_3)$ are known (unless we introduce $\theta_{1,0}$ which lacks a geometric interpretation),
we assume this to be the case. The discrete version of Bishop's frame is obtained by gauge transformation from the Frenet frame, by
demanding that
\[
\theta_{2,1} \ \to \ \theta_{2,1} + \Delta_1 - \Delta_2 = 0
\]
We can freely choose
\[
\Delta_1 \ = \ 0
\]
as an initial condition, and consequently we arrive at Bishop's frame by selecting 
\[
\Delta_2 \ = \ \theta_{2,1}
\]
For $\Delta_3$ we get similarly from
\[
\theta_{3,2} \ \to \ \theta_{3,2} + \Delta_2 - \Delta_3 \ = \ 0
\]
that 
\[
\Delta_3 \ = \ \theta_{2,1} + \theta_{3,2}
\]
and thus the discrete version of Bishop's parallel transport frame is related to the discrete Frenet frame by
gauge transformations
\[
\Delta_k \ = \ \sum_{i=1}^{k-1} \theta_{i+1,i}
\]
When we substitute this in (\ref{expRd}) with (\ref{thetaD}), we find that the transfer matrix (\ref{expRd}) simplifies into
\begin{equation}
\mathcal R^B_{i+1,i} = \left( \begin{matrix} 1+ \cos^2 \Theta_\Delta  ( \cos \psi-1)     \hskip 0.1cm &
\sin \Theta_\Delta \cos \Theta_\Delta ( \cos \psi - 1)  \hskip 0.1cm  & - \cos \Theta_\Delta \sin \psi 
\\
\sin \Theta_\Delta  \cos \Theta_\Delta  (\cos \psi - 1)    \hskip 0.1cm &
1+ \sin^2 \Theta_\Delta   (\cos \psi -1 )   \hskip 0.1cm  & - \sin \Theta_\Delta\sin \psi
 \\
\cos \Theta_\Delta \, \sin\psi & 
\sin\Theta_\Delta \, \sin\psi & \cos \psi 
\\
 \end{matrix} \right)_{\hskip -0.2cm i+1 , i} 
\la{expRd2}
\end{equation}
where now
\[
\Theta_\Delta \ \equiv \ \sum_{k=1}^{i} \theta_{k+1,k} 
\]
and with (\ref{DFE3}), we can  construct the discrete version of Bishop's parallel transport frame at each vertex $C_i$.

\section{ {\bf IV:} Framing of Folded Proteins}

As an application we  utilize the DF equation to investigate the
framing of the folded proteins in the Protein Data Bank (PDB) \cite{pdb}. We are  particularly interested in the existence and characterization of
a  {\it preferred} framing that  derives and directly reflects the physical  properties of the folded proteins.   
The identification  of such a preferred  framing, if it exists,  should help to pinpoint the 
physical principles that determine how  proteins fold.

From the PDB we get the three dimensional coordinates of all the different atoms in a folded protein. The overall fold  geometry  
is described by the location of the central $C_\alpha$ carbons that determine the protein backbone. We
take the $C_\alpha$ carbons  to be the vertices in a discrete and piecewise linear curve that models the backbone.
We then use the  $C_\alpha$ coordinates  to compute the corresponding Frenet framing. For this we  first apply (\ref{t}), (\ref{b}), (\ref{n}) 
to obtain the orthonormal basis vectors at each vertex. We then construct the transfer matrices by evaluating the bond and torsion angles  
from (\ref{bond}) and (\ref{tors}).

\subsection{{\bf A:}  ${\mathbb Z_2}$ Frenet framing and solitons}

We start by analyzing in detail an explicit example,  the chicken villin headpiece 
subdomain HP35 (PDB code 1YRF \cite{pdb}). This is a naturally existing 35-residue protein, with  three $\alpha$-helices 
separated from each other by two loops. This protein continues to be the subject to very extensive 
studies both experimentally \cite{knight}-\cite{wick} and theoretically \cite{pande}-\cite{fred}.  
We note that the overall resolution in the experimental x-ray data in PDB is 1.07\.A  in  RMSD \cite{chiu}.

We first compute the backbone Frenet frame bond and torsion angles ($\psi_{i+1,i}, \theta_{i+1,i}$) from  the 
PDB coordinates of the HP35 $C_\alpha$ carbons.   The result is shown in Figure 7 (left).
\begin{figure}[h]
        \centering
                \includegraphics[width=1.0\textwidth]{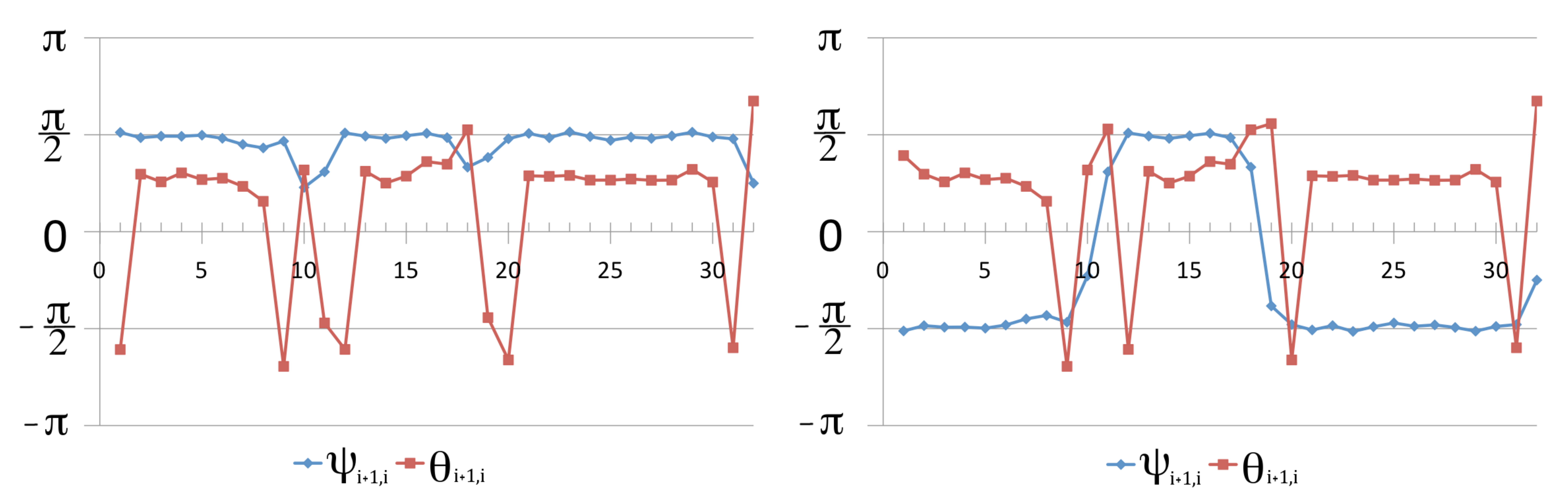}
        \caption{
      { {\bf Left:} The Frenet frame bond angle (blue) and torsion angle  (red) along the HP35 backbone. In this frame the potential presence of 
      an inflection point is only visible in large local variations of torsion angle. {\bf Right:} The outcome of 
      $\mathbb Z_2$ gauge transformations (\ref{dsgau}) at the loop regions.
      The result clearly reveals the presence of inflection points, they are located between
      the sites where the (gauge transformed) bond angle changes its sign. This can also be used to dentify the center of the loop. 
      Note how closely the profile of the
      bond angle in the right hand side  picture resembles that of  the kink-soliton in the {\it r.h.s.} of Figure 2.}
                }
       \label{Figure 7:}
\end{figure}

We inquire whether the loop regions contain inflection points.  As  we have previously explained  for example 
in connection of Figure 6,  the inflection points  can be difficult to identify in terms of the bond angles of the discrete
Frenet framing alone. But as apparent  from Figure 6, we can expect  that an inflection point  is located in the vicinity of vertices
where the Frenet  frame torsion angle is subject to strong local fluctuations. Thus we proceed to inspect the data in Figure 7 (left) 
using  the gauge transformation (\ref{dsgau}) to scrutinize the loop regions where the Frenet torsion angle in Figure 7 (left) 
is strongly fluctuating.  This leads  us to a particular version of the  $\mathbb Z_2$ Frenet frame, with bond and torsion angles as in
Figure 7 (right). 

By  comparing the bond angles in Figure 7 (right)  with the kink-soliton profile in the right hand side of Figure 2 we observe that the bond angles of
our gauge transformed frames at each of the loops have assumed the distinctive hallmark profile of  a (discrete) 
kink-soliton that interpolates between the adjacent $\alpha$-helices. In particular, we can unambiguously pinpoint the centers of the
loops to the locations of the inflection points on the curve: The inflection points are  between the vertices where the bond angle in our
gauge transformed frame changes its sign.

We have performed a similar analysis to several proteins  in the PDB, and some of our results where the techniques of
the present article are utilized have been reported in \cite{maxim}, \cite{nora}, \cite{xubiao}. The  results are remarkably consistent:  
In every secondary superstructure that we have studied 
where a loop  connects two $\alpha$-helices and/or $\beta$-strands,  after appropriate $\mathbb Z_2$
gauge transformations the profile of the bond angles
in the loop can be described with sub-\.Angstr\"om accuracy  in terms of a discrete version of the kink-soliton in Figure 2. The two asymptotic  ground states 
at $s = \pm c$ in this Figure correspond  to the $\alpha$-helices  and/or $\beta$-strands at the ends of the loop. For
the $\alpha$-helices we have the Frenet frame values very close to
\[
(\psi,\theta)_\alpha \  \approx \ (1.57, 0.87) \ \sim \ (\frac{\pi}{2}, 1) 
\]
The $\beta$ strands  can also be interpreted as helices, but in the "collapsed" limit with the approximative values
\[
(\psi, \theta)_\beta \ \approx \ (\pm 1.0 , - 2.9) \sim \ (\pm 1, -\pi )
\]
Consequently it appears  that these $\alpha$-helix/$\beta$-strand - loop - $\alpha$-helix/$\beta$-strand 
superstructures are indeed   {\it inflection point solitons} with the qualitative 
profile of (\ref{ys}).  We remark  that a long loop may also consist of a number of inflection points {\it i.e.} it can be a multi-soliton 
configuration.

\subsection{ {\bf B:} Physics based framing}

In every amino acid except glycine, there is a $C_\beta$ carbon that is covalently bonded to a $C_\alpha$ carbon.  
The positioning of these  $C_\beta$ carbons in relation to their  $C_\alpha$ carbons characterizes  the relative orientation of the 
amino acid side chains  along the protein backbone, and can be used to introduce a distinctive framing of the backbone; 
the case of glycine can be treated like that of an inflection point.  Since
the interactions between different amino acids are presumed to have a pivotal  r\^ole both during the folding process and
in the stabilization of the native fold, the  $C_\beta$ framing should  be a natural choice to  intimately reflect  the physical principles 
that determine the fold geometry of the backbone.
Consequently one way to try and understand the physical principles that determine  how a protein folds,  could be  to  investigate the
$C_\beta$ framing along  the protein backbone. 
Here we propose that a practical  approach  is to  look for  
gauge parameters (\ref{discso2}) that relate the $C_\beta$ frames to some purely geometrically determined frames such as the Frenet frames 
or parallel transport  frames. The identification of the rules that determine  the relevant gauge parameters $\Delta_i$ should then provide insight to the 
physical principles that underlie the protein folding phenomenon.

The $C_\beta$  framing is constructed from  the tangent vectors $\mathbf t$ of the backbone and  the 
unit vectors $\mathbf c$  that point from the $C_\alpha$ carbons towards their $C_\beta$ carbon. The framing is obtained by Gram-Schmidt algorithm,
by first introducing the unit vector 
\[
\mathbf p \ = \  \frac{ \mathbf t \times \mathbf c} { || \mathbf t \times \mathbf c ||}
\]
and then completing it into an orthonormal  frame $(\mathbf t, \mathbf p, \mathbf q)$ at each $C_\alpha$ vertex, where
\[
\mathbf  q\  = \  \mathbf t \times \mathbf p
\] 

In order to characterize the rules that determine the gauge parameter $\Delta_i$ relating a $C_\beta$ frame to the corresponding 
Frenet frame, we have investigated the statistical distribution of the $\mathbf c_i$ vectors in the  PDB proteins in the 
Frenet framing of the backbone.
For this we introduce, at each backbone vertex, the inclination angle $\chi_i\in [0,\pi] $  between the tangent vector $\mathbf t_i$  
and the corresponding vector $\mathbf c_i$, together with  the azimuthal angle $\varphi_i\in [-\pi,\pi] $  
between the normal vector $\mathbf n_i$ and the projection of $\mathbf c_i$ on the  $({\bf n}_i, {\bf b}_i)$ plane; see Figure 8.
\begin{figure}[h]
        \centering
                \includegraphics[width=0.5\textwidth]{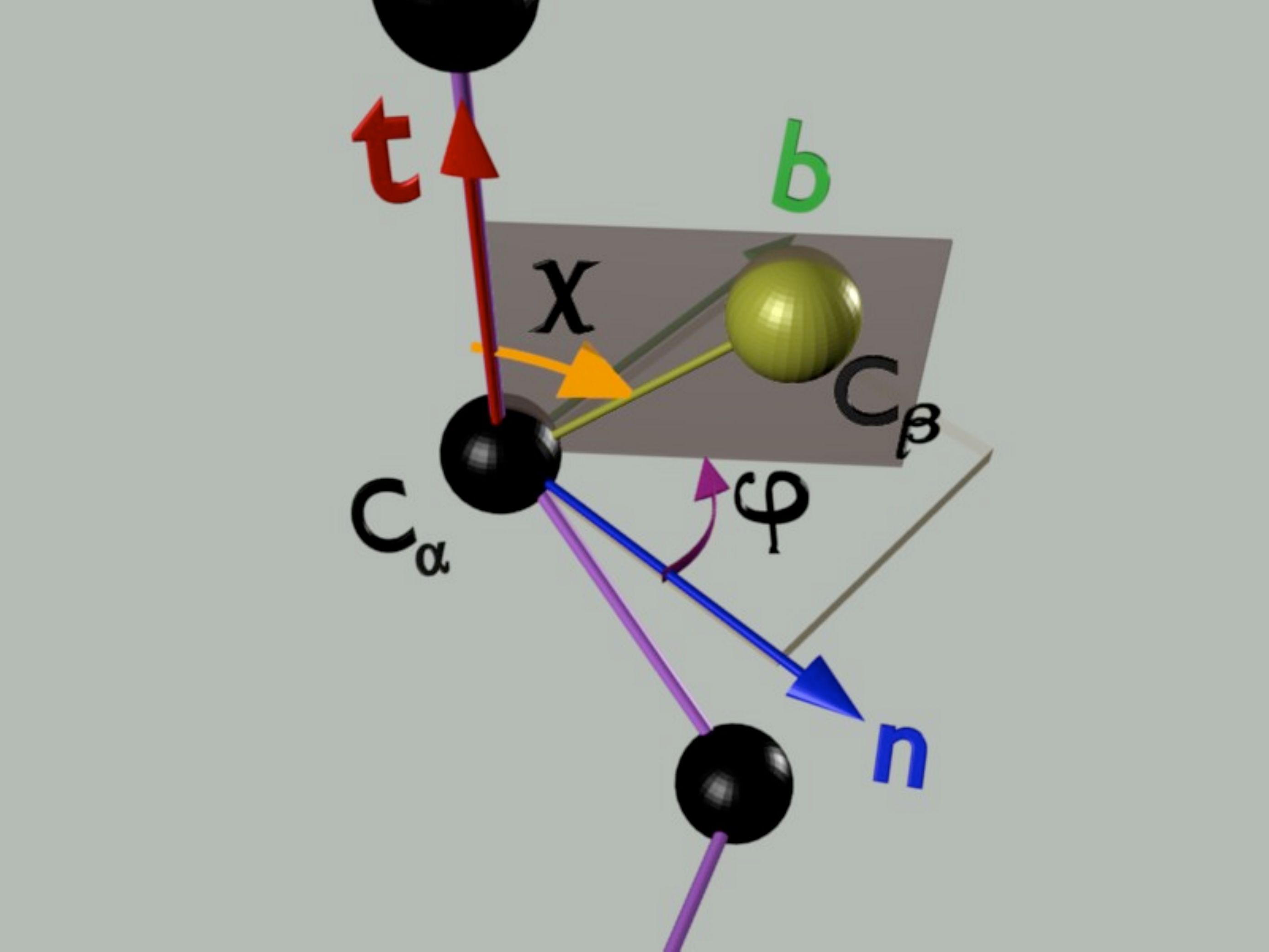}
        \caption{
      {The definition of the angles $\chi_i$ and $\varphi_i$ that  describe the location of the $i^{th}$ $C_\beta$ carbon
      with respect to the Frenet frame along the $C_\alpha$ backbone. The distance between the $C_\alpha$
      and $C_\beta$ carbons is within the range of 1.56-1.57 \.A. }
                }
       \label{Figure 8:}
\end{figure}

\vskip 0.3cm

We first consider the $C_\beta$ framing of the HP35. When we compute the directions of the individual vectors $\mathbf c_i$
in the Frenet frame, we get the result that we display in Figure 9. 
\begin{figure}[h]
        \centering
                \includegraphics[width=0.6\textwidth]{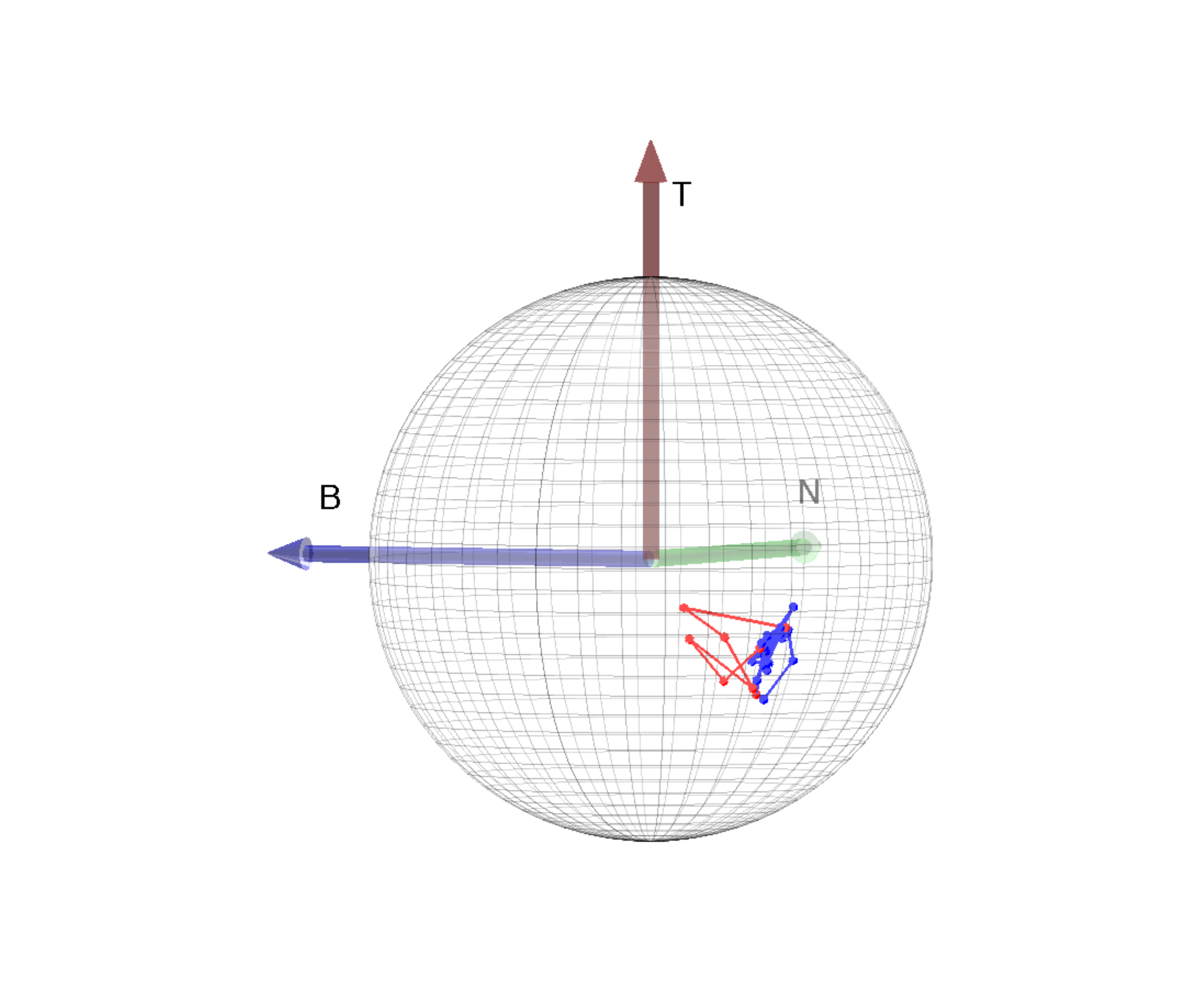}
        \caption{
      { The nutation in the direction of the vectors $\mathbf c_i$ in the Frenet frame along 1YRF backbone. The blue dots are the $C_\alpha$ carbons in the
     helices, and the red dots are the $C_\alpha$ carbons  that are located in the loops.  } }
       \label{Figure 9:}
\end{figure}
Remarkably, the directions of the $\mathbf c_i$ vectors in the Frenet frame
are {\it relatively } site independent. This implies that at least in the case of HP35, the parameters $\Delta_i$ that relate the $C_\beta$ frame to the
Frenet frame can be assigned to a high accuracy a constant and site independent value: The physically determined orthonormalized
$C_\beta$ frame appears to differ from the purely geometrically determined Frenet frame only by small nutations in the direction of the vectors $\mathbf c$ 
in the Frenet frame. We observe that these nutations are 
somewhat smaller in the helix regions that in the loops. 

We conclude that
since the Frenet framing of HP35 is determined entirely by the backbone geometry so are the orientations of the amino acids, with a 
surprisingly good accuracy. 

In the general case, we have inspected the correlation between the $C_\beta$ framing  and the Frenet framing by  performing a statistical analysis 
of the  directional distribution of the 
$\mathbf c$-vectors in the backbone  Frenet frames for all  amino acids in PDB.  Our results are summarized by
Figure 10 
\begin{figure}[h]
        \centering
                \includegraphics[width=0.8\textwidth]{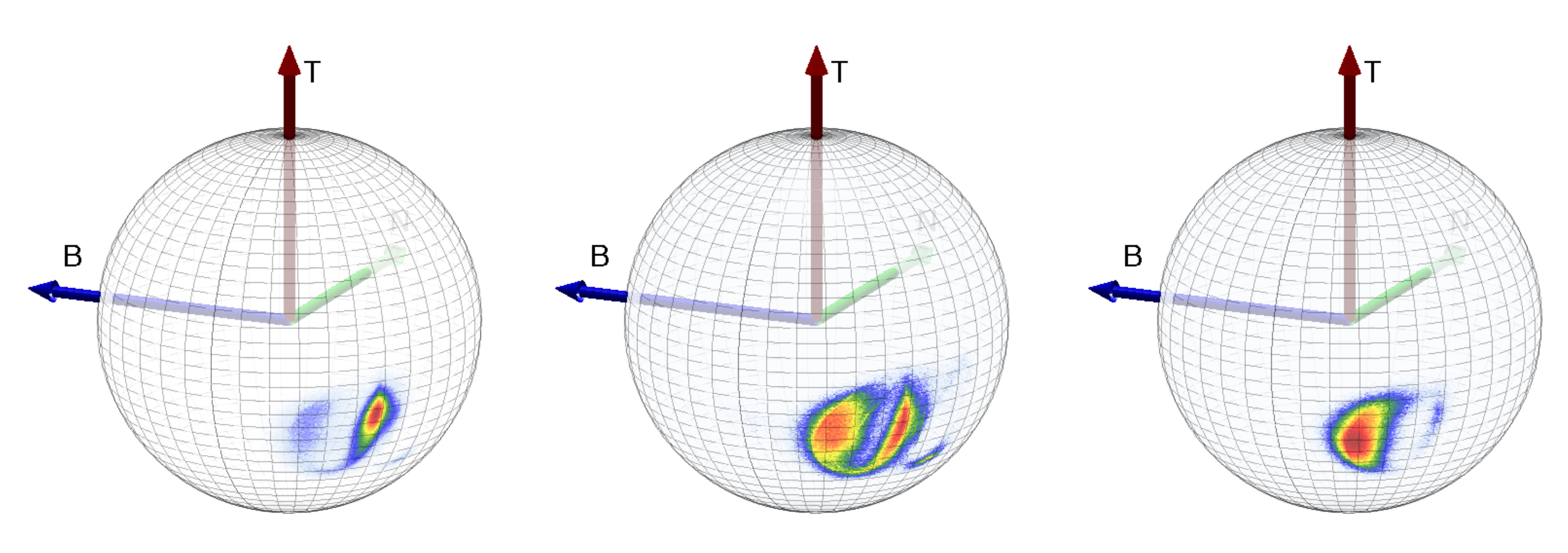}
        \caption{
      { Kent plots of the $C_\beta$ carbon vectors $\mathbf c$ for all sites of all proteins presently in PDB, with color intensity proportional to the
      number of vectors.   For  $\alpha$-helices {\it (left)}, 
      the direction of $\mathbf c$ nutates very little around the direction $(\chi,\varphi) \approx  (1.84, -2.20)$. For $\beta$-strands  {\it (right)} the nutation is somewhat
      more spread, but still very clearly concentrated around $(\chi,\varphi) \approx  (1.96,-2.47)$. 
      Finally, for loops {\it (center)} we observe the formation of a narrow arc 
      that connects the $\alpha$ and $\beta$ }
      directions.
                }
       \label{Figure 10:}
\end{figure}
where we display the statistical distribution of the angles $(\varphi_i, \chi_i)$   that we have defined in Figure 8. We have used the PDB 
definition to identify the three structures we display separately ($\alpha$-helix, $\beta$-strand, loop) but we note that there are sometimes
ambiguities in determining whether a particular amino acid belongs to a $\alpha$-helix,  $\beta$-strand or a loop in particular when the amino acid is 
located in the vicinity of the border between these three classes.

We find that the observation we have made in the
case of HP35 persists:  The orientations of the $C_\beta$ carbons in the Frenet frames are quite inert
and  essentially  protein and amino acid  independent. There is only a slight nutation around 
the statistical  average value. Furthermore, the directions for the $\alpha$-helices and $\beta$-strands are also almost the same,
the difference in the statistical average is surprisingly small but nevertheless noticeable. 
In the case of loops, we find that the  statistical distribution of the vector $\mathbf c$ in the Frenet frame
displays a thin band that connects the $\alpha$-helices and $\beta$-strands.  This universality is somewhat unexpected, since only
a small proportion of the loops connect an $\alpha$-helix with a $\beta$-strand. 

The overlapping regions between the three different classes 
in the Kent plots of Figure 10 can be at least partly explained by the uncertainty in classifying amino acids in the vicinity
of the border regions. We expect that a careful scrutiny of the class assignments of these amino acids 
will sharpen our statistical results. Alternatively, our approach could be developed into a technique 
to determine a more definite classification of those amino acids that are located in the border regions separating the $\alpha$-helices,
$\beta$-strands and the loops from each other.
But even at this level of classifying the amino acids  the results of our analysis  imply that almost independently of 
the protein, when we traverse its backbone by orienting 
the camera gaze direction
so that it remains fixed in the Frenet frames, the directions of the $C_\beta$ carbons are subject to only small nutations.

In Figure 11 we display the histograms for the components of the $C_\beta$ vectors $\mathbf c_i$ in terms of the $\chi$ and $\varphi$ angles
defined  in Figure 8. These histograms confirm that the directional variations of the $\mathbf c_i$ are surprisingly inert.
\begin{figure}[h]
        \centering
                \includegraphics[width=0.8\textwidth]{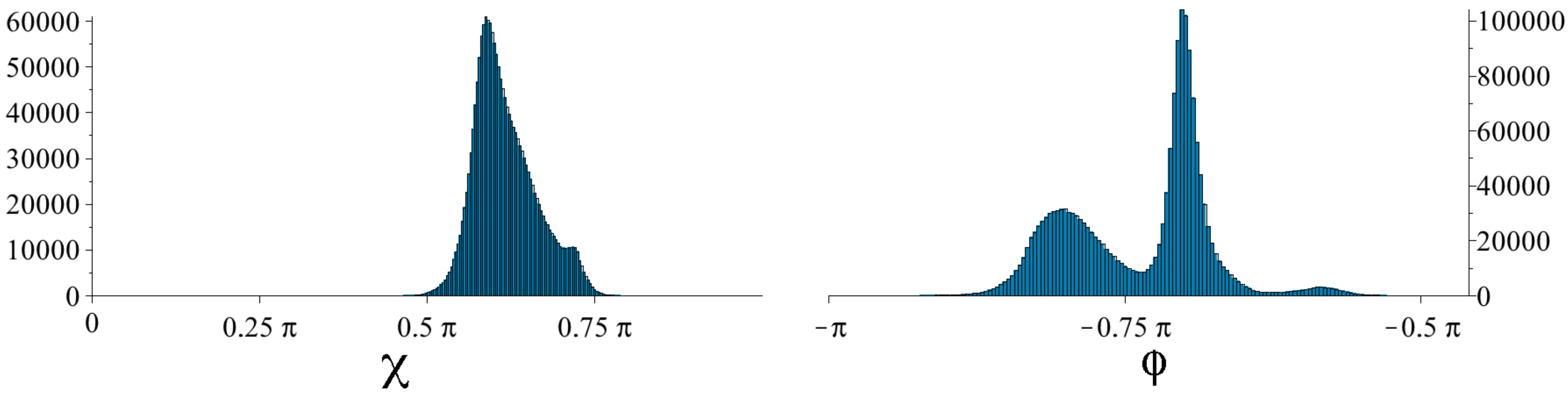}
        \caption{
      { Frenet frame histogram of the distribution of $(\chi,\varphi)$ angles displayed in Figure 10. for all $C_\beta$ in the PDB. The histogram shows 
      how the directions are subject to only very small deviations around their average values.  }
                }
       \label{Figure 11:}
\end{figure}

Finally, we have found that in  Bishop's  parallel transport frame 
the direction of the $C_\beta$ carbon does not lead to such a regular structure formation  as in the Frenet frame;  
See Figure 12 where we plot the statistical distributions of the vectors $\mathbf c$ in the Bishop's  frames. 
\begin{figure}[h]
        \centering
                \includegraphics[width=0.6\textwidth]{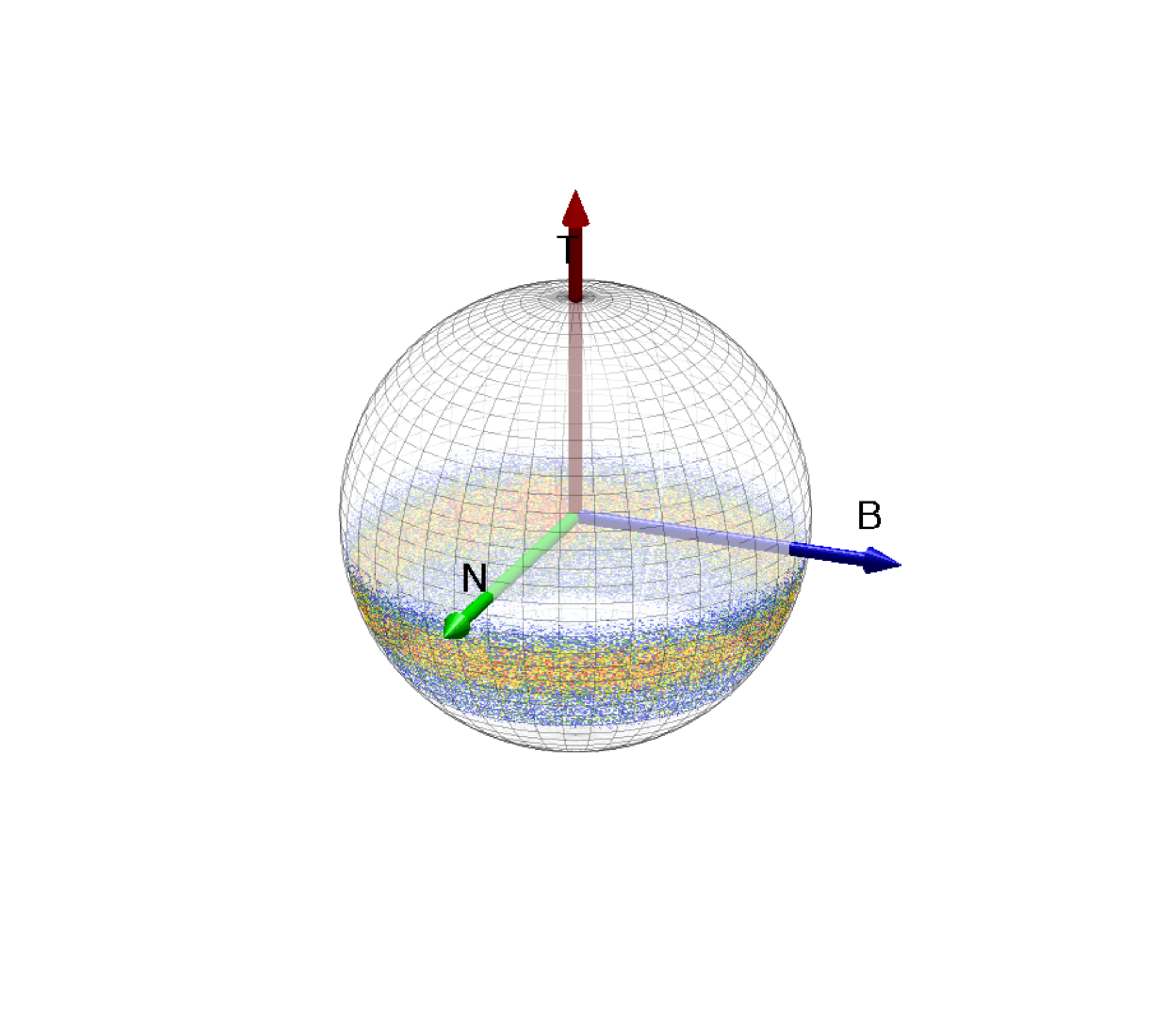}
        \caption{
      {The same as in Figure 10, but for all proteins in PDB using Bishop's parallel transport frame. In this frame the directions of the $C_\beta$ carbons are distributed in a  longitudinally uniform manner inside a segment of the Kent sphere. }
                }
       \label{Figure 12:}
\end{figure}

\section{ {\bf V:} Conclusions}

We have scrutinized the problem of frame determination along piecewise linear discrete curves, including those
with inflection points. Our approach is based on the transfer matrix method that has been previously  applied 
extensively to  investigate discrete integrable systems and lattice field theories.
The introduction of a  transfer matrix enables us to describe a framing in a covariant manner, with different frames related 
to each other by $SO(2)$ gauge transformations that correspond to rotations in the normal planes of the curve. 
In particular our construction  is not based on, and does not involve, any discretization of a continuum 
equation. Consequently  we can effortlessly describe
curves that become fractals in the limit where the lattice spacing  {\it a.k.a.} the length of line segments vanishes. But we have also verified that
if the continuum limit exists as a class $\mathcal C^3$ differentiable curve, we arrive at the  generalized version of   
the continuum Frenet equation. Furthermore, the manifest covariance of our formalism under frame rotations 
enables us to investigate the framing of a physically determined discrete curve in a manner where the framing 
is based on, and captures the properties of the underlying physical system. Consequently we expect 
that our formalism has wide  applications to the  visualization of discrete curves and 
the determination of camera gaze positions in a variety of scenarios.

One notable outcome of our analysis is the identification of inflection points with the centers of loops, and the interpretation
of loops as kink-solitons. In \cite{oma1}, \cite{oma2} we have already applied this identification to develop an Ansatz based on (\ref{ys}),
to succesfully describe the native folds of PDB proteins in terms of elementary functions.

As an example we have investigated the framing of folded proteins in the Protein Data Bank. In this case  no {\it valable} continuum description
exist, due to the fact that the universality class of folded proteins is characterized by the presence of a nontrivial Hausdorff dimension. Consequently any framing 
of folded proteins should be inherently discrete. In order to introduce a framing that  directly relates to the physical properties of a folded protein, 
we have employed the relative orientation of the $C_\beta$ carbons in the amino acids  with respect to the ensuing 
backbone central $C_\alpha$ carbons.  We have statistically analyzed the relative orientation of these $C_\beta$ frames to the 
geometrically determined Frenet frames of the PDB proteins. We have found that  the two framings are almost identical,  they
differ from each other  only by a practically amino acid independent global frame rotation: 
For the $\alpha$-helices the nutation in the orientation of the $C_\beta$ carbons in the Frenet frame is {\it very} sharply concentrated around
its statistically determined average direction. For $\beta$-strands the result is very similar, with only a relatively small increase in the amplitude
of nutations.   Finally, in the case of loops we find that the orientation of the $C_\beta$ carbons 
oscillates along a narrow circular arc that connects the $\alpha$-helices and $\beta$-carbons. 
In each case we have used the definition employed in the Protein Data Bank to identify  the helix/loop class of the amino acid,  and we note that the existing
criteria for determining this class in the case of  an  amino acid that is located in the vicinity of the terminals of each structure
is subject to interpretations. Consequently we propose that there are several border line cases that interfere destructively 
with the accuracy of our statistically determined results. We hope that our framing technique will eventually provide a refinement of the existing classification 
principles.  The biophysical interpretation and biological relevance of our observations will be reported elsewhere.

\section{ Acknowledgements} We thank Maxim Chernodub for many valuable discussions and Jack Quine for communications.


\end{document}